 \documentclass[11pt,a4paper]{article}

\usepackage{ifthen}
\newboolean{full}
\setboolean{full}{true}

\usepackage{epsfig}
\usepackage{amsfonts,amssymb,amstext,amsmath}
\usepackage{shadow,times}
 \usepackage{graphicx}
\usepackage{algorithm,algorithmic}
\usepackage{color}
\usepackage{floatflt}
\usepackage{xspace}
\usepackage{bm}
\usepackage{ifpdf}
\usepackage{picins}

\ifthenelse{\boolean{full}}{ %% full version exclusives

  \usepackage{fullpage}
  \usepackage[letterpaper,margin=1.1in]{geometry}

  \newtheorem{theorem}{Theorem}
  \newtheorem{lemma}[theorem]{Lemma}
  \newtheorem{corollary}[theorem]{Corollary}
  \newtheorem{definition}[theorem]{Definition}

  \newcommand{\email}[1]{#1}
  \newcounter{note}[section]

}{}

\newcommand\ignore[1]{}
\def\tab{\hspace{5mm}}

\newenvironment{pf}{\noindent{\bf Proof:  }}{\hfill\rule{2mm}{2mm}\medskip}
\def\np{$\mathcal{NP}$}
\def\sse{\subseteq}

\def\bnd{{\sf Bnd}}
\def\inc{{\sf Inc}}
\def\m{\mathcal{M}}
\def\bs{\mathcal{B}}
\def\cs{\mathcal{C}}
\def\n{\mathcal{N}}
\def\lam{\mathcal{L}}
\def\s{{\sf Sp}}
\def\sse{\subseteq}
\allowdisplaybreaks
\newcommand{\lplatt}{\ensuremath{\mathcal{LP}_{\mbox{lat}}}}

\newcommand{\NP}{\ensuremath{\text{NP}}}

\newcommand{\is}{\ensuremath{\mathcal{I}}\xspace}

\newcommand{\rank}{\ensuremath{\sf rank}}

\newcommand{\F}{\ensuremath{\mathcal{F}}}
\newcommand{\f}{\ensuremath{\mathcal{F}}}
\newcommand{\T}{\ensuremath{\mathcal{T}}}
\newcommand{\B}{\ensuremath{\mathcal{B}}}
\newcommand{\R}{\mathbb{R}}

\newcommand{\local}{\mathrm{local}}

\newtheorem{cl}[theorem]{Claim}

\title{On Generalizations of Network Design Problems \\with Degree Bounds}

\ifthenelse{\boolean{full}}{
 \author{
    Nikhil Bansal \thanks{
      IBM T.J. Watson Research Center. \email{\{nikhil,rohitk,viswanath\}@us.ibm.com}}
    \and
    Rohit Khandekar~$^*$
    \and
    Jochen K{\"o}nemann\thanks{
      University of Waterloo. \email{jochen@uwaterloo.ca}}
    \and
    Viswanath Nagarajan~$^*$
    \and
    Britta Peis\thanks{
      Technische Universit\"at Berlin. \email{peis@math.tu-berlin.de}}
  }
}{
  \author{
    Nikhil Bansal~\inst{1}
    \and
    Rohit Khandekar~\inst{1}
    \and
    Jochen K{\"o}nemann~\inst{2}
    \and
    \\Viswanath Nagarajan~\inst{1}
    \and
    Britta Peis~\inst{3}}

  \institute{
    IBM T.J. Watson Research Center, Yorktown Heights, NY 10598, USA. \and
    University of Waterloo. \and Technische Universit\"at Berlin.
  }
}

\date{}

\begin{document}

\maketitle

\ignore{ text-abstract: Iterative rounding and relaxation have arguably become the method of choice in dealing with
unconstrained and constrained network design  problems. In this paper we extend the scope of the iterative relaxation
method in two directions: (1) by handling more complex degree constraints in the minimum spanning tree problem (namely,
laminar crossing spanning tree), and (2) by incorporating `degree bounds' in other combinatorial optimization problems
such as matroid intersection and lattice polyhedra. We give new or improved approximation algorithms, hardness results,
and integrality gaps for these problems. }

\begin{abstract}
  Iterative rounding and relaxation have arguably become the method of
  choice in dealing with unconstrained and constrained network design
  problems.
  % Starting with Jain's elegant iterative rounding scheme for
  % the generalized Steiner network problem, the technique has more
  % recently lead to breakthrough results in the area of constrained
  % network design, where a number of linear constraints are added to a
  % classical network design problem. Most notable among recent
  % successes is Singh \& Lau's +1-algorithm [STOC, 2007] for the
  % minimum-cost degree-bounded spanning tree problem.
  In this paper we
  extend the scope of the iterative relaxation method in two
  directions: (1) by handling more complex degree constraints in the
  minimum spanning tree problem (namely {\em laminar} crossing spanning tree), and (2) by incorporating `degree
  bounds' in other combinatorial optimization problems such as {\em
    matroid intersection} and {\em lattice polyhedra}.
We give new or improved approximation algorithms, hardness results, and integrality gaps for these problems.

\ifthenelse{\boolean{full}}{
  \begin{itemize}
  \item Our main result is a $(1,b+O(\log n))$-approximation algorithm
    for the {\em minimum crossing spanning tree} (MCST) problem with
    {\em laminar} degree constraints. The laminar MCST problem is a
    natural generalization of the well-studied bounded-degree MST, and
    is a special case of general crossing spanning tree.
%~[Bilo et    al. APPROX 2004].
We also give an additive $\Omega(\log^\alpha m)$
    hardness of approximation for {\em general} MCST, even in the
    absence of costs ($\alpha>0$ is a fixed constant, and $m$ is the
    number of degree constraints).

  \item We then consider the {\em crossing contra-polymatroid intersection}
    problem and obtain a $(2,2b+\Delta-1)$-approxi\-mation algorithm,
    where $\Delta$ is the maximum element frequency. This models for example
    the degree-bounded spanning-set intersection
    in two matroids.
    Finally, we introduce
    the {\em crossing lattice polyhedra} problem, and obtain a
    $(1,b+2\Delta-1)$ approximation under certain condition. This
    result provides a unified framework and common generalization of
    various problems studied previously, such as degree bounded
    matroids.
% (Kir\'aly et al., [IPCO, 2008]).
  \end{itemize}}{}

\end{abstract}

\section{Introduction}

Iterative rounding and relaxation have arguably become the method of choice in dealing with unconstrained and
constrained network design problems. Starting with Jain's elegant {\em iterative rounding} scheme for the generalized
Steiner network problem in \cite{J01}, an extension of this technique (iterative {\em relaxation}) has more recently
lead to breakthrough results in the area of constrained network design, where a number of linear constraints are added
to a classical network design problem. Such constraints arise naturally in a wide variety of practical applications,
and model limitations in processing power, bandwidth or budget. The design of powerful techniques to deal with these
problems is therefore an important goal.

The most widely studied constrained network design problem is the {\em
  minimum-cost degree-bounded spanning tree} problem. In an instance
of this problem, we are given an undirected graph, non-negative costs for the edges, and positive, integral
degree-bounds for each of the nodes.  The problem is easily seen to be \NP-hard, even in the absence of edge-costs,
since finding a spanning tree with maximum degree two is equivalent to finding a Hamiltonian Path. A variety of
techniques have been applied to this problem \cite{C+05,C+06,G06,KR02,KR05,R+93,RS06}, culminating in Singh and Lau's
breakthrough result in \cite{SL07}. They presented an algorithm that computes a spanning tree of at most optimum cost
whose degree at each vertex $v$ exceeds its bound by at most $1$, using the {\em
  iterative relaxation} framework developed in~\cite{LNSS07,SL07}.

The iterative relaxation technique has been applied to several constrained network design problems: spanning
tree~\cite{SL07}, survivable network design~\cite{LNSS07,LS08}, directed graphs with intersecting and crossing
super-modular connectivity~\cite{LNSS07,BKN08}. It has also been applied to degree bounded versions of matroids and
submodular flow~\cite{KLS08}.

In this paper we further extend the applicability of iterative relaxation, and obtain new or improved bicriteria
approximation results for minimum crossing spanning tree (MCST), crossing contra-polymatroid intersection, and crossing
lattice polyhedra.  We also provide some hardness results and integrality gaps for these problems.

\noindent {\bf Notation.} As is usual, when dealing with an undirected graph $G=(V,E)$, for any $S\sse V$ we let
$\delta_G (S):=\{(u,v)\in E\mid u\in S,~v\not\in S\}$. When the graph is clear from context, the subscript is dropped.
A collection $\{U_1,\cdots,U_t\}$ of vertex-sets is called {\em laminar} if for every pair $U_i,U_j$ in this
collection, we have $U_i\sse U_j$, $U_j\sse U_i$, or $U_i\cap U_j=\emptyset$. A $(\rho,f(b))$ approximation for minimum
cost degree bounded problems refers to a solution that (1) has cost at most $\rho$ times the optimum that satisfies the
degree bounds, and (2) satisfies the relaxed degree constraints in which a bound $b$ is replaced with a bound $f(b)$.

\subsection{Our Results, Techniques and Paper Outline}

\paragraph{Laminar MCST.}
Our main result is for a natural generalization of bounded-degree MST (called Laminar Minimum Crossing Spanning Tree or
{\em laminar MCST}), where we are given an edge-weighted undirected graph with a laminar family ${\cal L} =
\{S_i\}_{i=1}^m$ of vertex-sets having bounds $\{b_i\}_{i=1}^m$; and the goal is to compute a spanning tree of minimum
cost that contains at most $b_i$ edges from $\delta(S_i)$ for each $i\in[m]$.

The motivation behind this problem is in designing a network where there is a hierarchy (i.e. laminar family) of
service providers that control nodes (i.e. vertices). The number of edges crossing the boundary of any service provider
(i.e. its vertex-cut) represents some cost to this provider, and is therefore limited. The laminar MCST problem
precisely models the question of connecting all nodes in the network while satisfying bounds imposed by all the service
providers.

From a theoretical viewpoint, cut systems induced by laminar families are well studied, and are known to display rich
structure. For example, {\em one-way cut-incidence matrices} are matrices whose rows are incidence vectors of directed
cuts induced by the vertex-sets of a laminar family; It is well known (e.g., see \cite{KV08}) that such matrices are
totally unimodular. Using the laminar structure of degree-constraints and the iterative relaxation framework, we obtain
the following main result, and present its proof in Section \ref{sec:cpt}.

\begin{theorem}
  \label{thm:lam} There is a polynomial time $(1,b+O(\log n))$
  bicriteria approximation algorithm for laminar MCST. That is, the
  cost is no more than the optimum cost and the degree violation is at
  most additive $O(\log n)$. This guarantee is relative to the natural
  LP relaxation.
\end{theorem}

This guarantee is substantially stronger than what follows from known results for the general {\em minimum crossing
spanning tree} (MCST) problem: where the degree bounds could be on arbitrary edge-subsets $E_1,\ldots,E_m$.  In
particular, for general MCST a $(1,b+\Delta-1)$~\cite{BKN08,KLS08} is known where $\Delta$ is the maximum number of
degree-bounds an edge appears in. However, this guarantee is not useful for laminar MCST as $\Delta$ can be as large as
$\Omega(n)$ in this case. If a multiplicative factor in the degree violation is allowed, Chekuri et al.~\cite{CVZ09}
recently gave a very elegant $\left(1,(1+\epsilon)b+O(\frac1\epsilon \log m)\right)$ guarantee (which subsumes the
previous best $(O(\log n), O(\log m)\, b)$~\cite{B+04} result). However, these results also cannot be used to obtain a
small additive violation, especially if $b$ is large. In particular, both the results \cite{B+04,CVZ09} for general
MCST are based on the natural LP relaxation, for which there is an integrality gap of $b+\Omega(\sqrt{n})$ even without
regard to costs and when $m=O(n)$~\cite{S08} \ifthenelse{\boolean{full}}{(see also Section \ref{sec:mcst-gap})}{(see
  also \cite{BK+10})}. On the other hand,
Theorem~\ref{thm:lam} shows that a purely additive $O(\log n)$ guarantee on degree (relative to the LP relaxation and
even in presence of costs) is indeed achievable for MCST, when the degree-bounds arise from a laminar cut-family.

The algorithm in Theorem~\ref{thm:lam} is based on iterative relaxation and uses two main new ideas. Firstly, we drop a
carefully chosen {\em constant fraction of degree-constraints} in each iteration. This is crucial as it can be shown
that dropping one constraint at a time as in the usual applications of iterative relaxation can indeed lead to a degree
violation of $\Omega(\Delta)$. Secondly, the algorithm does not just drop degree constraints, but in some iterations it
also {\em generates new degree
  constraints}, by merging existing degree constraints.

All previous applications of iterative relaxation to constrained network design treat connectivity and degree
constraints rather asymmetrically. While the structure of the connectivity constraints of the underlying LP is used
crucially (e.g., in the ubiquitous uncrossing argument), the handling of degree constraints is remarkably simple.
Constraints are dropped one by one, and the final performance of the algorithm is good only if the number of side
constraints is small (e.g., in recent work by Grandoni et al.~\cite{GRS09}), or if their structure is simple (e.g., if
the `frequency' of each element is small). In contrast, our algorithm for laminar MCST exploits the structure of degree
constraints in a non-trivial manner.

\paragraph{Hardness Results.}

We obtain the following hardness of approximation for the {\em general MCST} problem (and its matroid counterpart). In
particular this rules out any algorithm for MCST that has additive constant degree violation, even without regard to
costs.

\begin{theorem}\label{th:mcst-hard}
  Unless \np~has quasi-polynomial time algorithms, the MCST problem
  admits no polynomial time $O(\log^\alpha m)$ additive approximation for the
  degree bounds for some
  constant $\alpha>0$; this holds even when there are no costs.
\end{theorem}

The proof for this theorem is given in Section \ref{sec:hard}, and uses a a two-step reduction from the well-known {\em
Label Cover} problem. First, we show hardness for a {\em uniform} matroid instance. In a second step, we then
demonstrate how this implies the result for MCST claimed in Theorem \ref{th:mcst-hard}.

Note that our hardness bound nearly matches the result obtained by Chekuri et al. in \cite{CVZ09}. We note however that
in terms of {\em
  purely} additive degree guarantees, a large gap remains.  As noted
above, there is a much stronger lower bound of $b+\Omega(\sqrt{n})$ for LP-based algorithms~\cite{S08} (even without
regard to costs), which is based on discrepancy.  In light of the small number of known hardness results for
discrepancy type problems, it is unclear how our bounds for MCST could be strengthened.

\ifthenelse{\boolean{full}}{
  An interesting consequence of the hardness result in
  Theorem~\ref{th:mcst-hard} is for the {\em robust (or min-max)
    $k$-median} problem~\cite{AGGN08}. In this problem, there are $m$
  different client-sets in a metric and the goal is to open $k$
  facilities that are simultaneously good (in terms of the $k$-median
  objective) for all the client-sets.  Anthony et al.~\cite{AGGN08}
  obtained a logarithmic approximation algorithm for this problem, and
  showed that it is hard to approximate better than factor $2$. The
  following result
  shows that the robust $k$-median problem is indeed
  harder to approximate than usual $k$-median, for which
  $O(1)$-approximations are known~\cite{CGTS02,A+01}.
  We present its proof in Section \ref{sec:kmed-hard}.

  \begin{corollary}\label{cor:k-med-hard}
    Robust $k$-median is $\Omega(\log^\alpha m)$-hard to approximate
    even on uniform metrics (for some fixed constant $\alpha>0$),
    assuming \np~does not have quasi-polynomial time algorithms.
  \end{corollary}
}{}

\paragraph{Degree Bounds in More General Settings.}
We consider crossing versions of other classic combinatorial optimization problems, namely {\em contra-polymatroid
intersection} and {\em  lattice polyhedra}~\cite{Sch03}. \ifthenelse{\boolean{full}}{}{
  We discuss our results briefly and defer the proofs to the full version of the paper~\cite{BK+10}.
}

\begin{definition}[Minimum crossing  contra-polymatroid intersection
  problem]\label{def.BoundedPolymatroidIntersection}
  Let $r_1,r_2:2^E\to \mathbb{Z}$ be two supermodular functions,
  $c:E\to \mathbb{R}_+$ and $\{E_i\}_{i\in I}$ be a collection of
  subsets of $E$ with corresponding bounds $\{b_i\}_{i\in I}$. Then
  the goal is to minimize:
  \begin{align*}
    \{ c^Tx ~\big| \quad & x(S)\ge \max\{r_1(S), r_2(S)\},
       \forall~ S\subseteq E; \\  & x(E_i)\le b_i,~~\forall~ i\in I;
      \quad x \in \{0,1\}^E \}.
  \end{align*}

\ignore{\begin{align*}
 \min\ &c^Tx\\
  &x(S)\ge \max\{r_1(S), r_2(S)\} &\forall S\subseteq E\\
  &x(E_i)\le b_i&\forall i\in [m]\\
  &x_e\in \{0,1\} &\forall e\in E.
\end{align*}}
\end{definition}

In particular, this definition captures the degree-bounded version of spanning-set intersection in two matroids (for
eg. the {\em bipartite edge-cover} problem). We note that this definition does not capture alternate notions of matroid
intersection, such as intersection of bases in two matroids; hence it does not apply to the degree-bounded arborescence
problem.~\footnote{In an earlier version of the paper~\cite{BK+10}, we had incorrectly claimed that our result extends
to degree-bounded arborescence.}

  Let $\Delta = \max_{e\in E} |\{ i\in [m] \mid e \in E_i\}|$ be the largest number of sets $E_i$
  that any element of $E$ belongs to, and refer to it as {\em frequency}.
\ifthenelse{\boolean{full}}{ The proof of this theorem can be found in
  Section \ref{sec:mcpi}.  }{}

\begin{theorem}\label{t.matroidintersection}
  Any optimal basic solution $x^*$ of the linear relaxation of the
  minimum crossing contra-polymatroid intersection problem can be rounded into an
  integral solution $\hat{x}$ such that:
$$\hat{x}(S)\ge \max\{r_1(S),  r_2(S)\}, \,\, \forall S\subseteq E; \qquad \hat{x}(E_i)\le 2b_i + \Delta -1, \,\, \forall i\in I; \quad\mbox{and}\quad c^T\hat{x}\le 2c^Tx^*.$$
\end{theorem}

The algorithm for this theorem again uses iterative relaxation, and its proof is based on a `fractional token' counting
argument similar to the one used in~\cite{BKN08}. We also observe that the natural iterative relaxation steps are
insufficient to obtain a better approximation guarantee.\\

\noindent {\bf Crossing Lattice Polyhedra.} Classical {\em lattice
  polyhedra} form a unified framework for various discrete optimization
problems and go back to Hoffman and Schwartz~\cite{HS78} who proved their integrality. They are polyhedra of type
$$\{x\in [0,1]^E\mid x(\rho(S))\ge r(S),\quad \forall S\in \mathcal{F}\}$$
where $\F$ is a \emph{consecutive submodular} lattice, $\rho: \F\rightarrow 2^E$ is a mapping from $\F$ to subsets of
the ground-set $E$, and $r\in \R^{\F}$ is supermodular.  A key property of lattice polyhedra is that the uncrossing
technique can be applied which turns out to be crucial in almost all iterative relaxation approaches for optimization
problems with degree bounds.  We refer the reader to \cite{Sch03} for a more comprehensive treatment of this subject.

We generalize our work further to \emph{crossing
  lattice polyhedra} which arise from classical lattice polyhedra by
adding ``degree-constraints" of the form $a_i\le x(E_i)\le b_i$ for a given collection $\{E_i\subseteq E\mid i\in I\}$
and lower and upper bounds $a,b\in \R^I$. \ifthenelse{\boolean{full}}{
  We mention two (of several) examples which are
  covered by this model:

  \smallskip \emph{Example 1: Crossing matroid basis.} Here $\F=2^E$,
  $\rho$ is the identity map, and the partial order in $\F$ is the canonical one that is induced by set inclusion.
  Function $r: 2^E \rightarrow \mathbb{N}$ is
  defined as $r(S) =\rank(V) - \rank(V\setminus S)$; where $E$ is the
  ground-set of the matroid and \rank~ is its {\em rank function}. The
  crossing matroid basis problem finds the minimum cost basis in the
  matroid satisfying degree bounds.

  \smallskip \emph{Example 2: Crossing planar min cut.} Let $G=(V,E)$ be
  a (directed or undirected) $s,t$-planar graph (along with an embedding) with
  $s,t\in V$. Here elements of $\F$ correspond to $s$-$t$ paths in $G$
  ($\rho$ maps each element of $\F$ to the edge-set of that $s-t$ path),
  and the partial order in $\F$ relates paths where one is below/above
  the other in the planar embedding of $G$. The rank function is the
  constant all-ones function. The crossing planar min-cut problem
  involves finding a minimum cost $s-t$ cut in $G$ that obeys the degree
  bounds.
}{
  We mention that this model covers several important
  applications including the crossing matroid basis and crossing
  planar mincut problems, among others.
}

We can show that the standard LP relaxation for the general crossing lattice polyhedron problem is weak;
  in Section \ref{sec:lph-gap} we give instances of crossing planar min-cut
  (i.e., Example~2 above) where the LP-relaxation is feasible, but any
  integral solution violates some degree-bound by
  $\Omega(\sqrt{n})$.
For this reason, we henceforth focus on a restricted class of crossing lattice polyhedra in which the underlying
lattice $(\F,\le)$ satisfies the following monotonicity property

$$(*) \quad S<T \implies |\rho(S)|<|\rho(T)| \qquad\forall~ S,T\in \F.$$

We obtain the following theorem whose proof is given in \ifthenelse{\boolean{full}}{
  Section \ref{sec:mclp}.
}{
  \cite{BK+10}.
}

\begin{theorem}\label{t.BoundedLatticePolyhedra}
  For any instance of the crossing lattice polyhedron problem in which
  $\F$ satisfies property $(*)$, there exists an algorithm that
  computes an integral solution of cost at most the optimal, where all
  rank constraints are satisfied, and each degree bound is violated by
  at most an additive $2\Delta-1$.
\end{theorem}

We note that the above property $(*)$ is satisfied for matroids, and hence Theorem~\ref{t.BoundedLatticePolyhedra}
matches the previously best-known bound~\cite{KLS08} for degree bounded matroids (with both upper/lower bounds). Also
note that property $(*)$ holds whenever $\F$ is ordered by inclusion. In this special case, we can improve the result
to an additive $\Delta-1$ approximation if only upper bounds are given.

\subsection{Related Work}

As mentioned earlier, the basic bounded-degree MST problem has been extensively
studied~\cite{C+05,C+06,G06,KR02,KR05,R+93,RS06,SL07}. The iterative relaxation technique for degree-con\-strained
problems was developed in~\cite{LNSS07,SL07}.

%MCST papers
MCST was first introduced by Bilo et al.~\cite{B+04}, who presented a randomized-rounding algorithm that computes a
tree of cost $O(\log n)$ times the optimum where each degree constraint is violated by a multiplicative $O(\log n)$
factor and an additive $O(\log m)$ term. Subsequently, Bansal et al.~\cite{BKN08} gave an algorithm that attains an
optimal cost guarantee and an additive $\Delta-1$ guarantee on degree; recall that $\Delta$ is the maximum number of
degree constraints that an edge lies in. This algorithm used iterative relaxation as its main tool. Recently, Chekuri
et al.~\cite{CVZ09} obtained an improved $\left(1,(1+\epsilon)b +O(\frac1\epsilon \log
  m)\right)$ approximation algorithm for MCST, for any $\epsilon>0$;
this algorithm is based on pipage rounding.

%Matroid basis paper
The minimum crossing matroid basis problem was introduced in~\cite{KLS08}, where the authors used iterative relaxation
to obtain (1) $(1,b+\Delta-1)$-approximation when there are only upper bounds on degree, and (2)
$(1,b+2\Delta-1)$-approximation in the presence of both upper and lowed degree-bounds. The~\cite{CVZ09} result also
holds in this matroid setting.~\cite{KLS08} also considered a degree-bounded version of the {\em submodular flow}
problem and gave a $(1,b+1)$ approximation guarantee.

%bnd-deg arb papers
The bounded-degree arborescence problem was considered in Lau et al.~\cite{LNSS07}, where a $(2,2b+2)$ approximation
guarantee was obtained. Subsequently Bansal et al.~\cite{BKN08} designed an algorithm that for any $0 < \epsilon\le
1/2$, achieves a $(1/\epsilon, b_v/(1-\epsilon) + 4)$ approximation guarantee. They also showed that this guarantee is
the best one can hope for via the natural LP relaxation (for every $0 < \epsilon\le 1/2$). In the absence of
edge-costs, ~\cite{BKN08} gave an algorithm that violates degree bounds by at most an additive two. Recently
Nutov~\cite{N08} studied the arborescence problem under {\em weighted} degree constraints, and gave a $(2,5b)$
approximation for it.
%; ~\cite{N08} also gave another proof of the $(2,2b+2)$ approximation

%lattice polyhedra
Lattice polyhedra were first investigated by Hoffman and Schwartz~\cite{HS78} and the natural LP relaxation was shown
to be totally dual integral.  Even though greedy-type algorithms are known for all examples mentioned earlier, so far
no combinatorial algorithm has been found for lattice polyhedra in general. Two-phase greedy algorithms have been
established only in cases where an underlying rank function satisfies a monotonicity property
~\cite{Fra99},~\cite{FP08}.

\section{Crossing Spanning Tree with Laminar degree bounds}
\label{sec:cpt}

In this section we prove Theorem~\ref{thm:lam} by presenting an iterative relaxation-based algorithm with the stated
performance guarantee.  During its execution, the algorithm selects and deletes edges, and it modifies the given
laminar family of degree bounds. A generic iteration starts with a subset $F$ of edges already picked in the solution,
a subset $E$ of {\em undecided} edges, i.e., the edges not yet picked or dropped from the solution, a laminar family
$\lam$ on $V$, and residual degree bounds $b(S)$ for each $S\in\lam$.

The laminar family $\lam$ has a natural forest-like structure with {\em nodes} corresponding to each element of $\lam$.
A node $S\in\lam$ is called the {\em parent} of node $C\in\lam$ if $S$ is the inclusion-wise minimal set in
$\lam\setminus\{C\}$ that contains $C$; and $C$ is called a {\em child} of $S$. Node $D\in\lam$ is called a {\em
grandchild} of node $S\in\lam$ if $S$ is the parent of $D$'s parent. Nodes $S,T\in\lam$ are {\em siblings} if they have
the same parent node. A node that has no parent is called {\em
  root}. The {\em level} of any node $S\in\lam$ is the length of the
path in this forest from $S$ to the root of its tree. We also maintain a {\em linear ordering} of the children of each
$\lam$-node. A subset $\bs\sse \lam$ is called {\em consecutive} if all nodes in $\bs$ are siblings (with parent $S$)
and they appear consecutively in the ordering of $S$'s children. In any iteration $(F,E,\lam,b)$, the algorithm solves
the following LP relaxation of the residual problem. \ignore{Here $E(S)$ denotes the edges in $E$ induced on
  $S\subseteq V$ while $\delta(S)$ denotes the edges in $E$ crossing
  $S\subseteq V$.}

%\vspace{-4mm}
\begin{small}\begin{align}
\min \quad & \sum_{e \in E} c_e x_e \label{lp} \\
\mbox{s.t.} \quad &  x(E(V)) = |V|-|F|-1 \notag \\
& x(E(U)) \leq |U| - |F(U)| - 1 \quad & & \forall U \subset V \notag\\
& x(\delta_E(S)) \leq b(S) & & \forall S \in \lam \notag \\
& x_e \geq 0 \quad & & \forall e\in E \notag
\end{align}
\end{small}
%\vspace{-4mm}

For any vertex-subset $W\sse V$ and edge-set $H$, we let $H(W):=\{(u,v)\in H\mid u,v\in W\}$ denote the edges induced
on $W$; and $\delta_H(W):= \{(u,v)\in H\mid u\in W,~v\not\in W\}$ the set of edges crossing $W$. The first two sets of
constraints are spanning tree constraints while the third set corresponds to the degree bounds. Let $x$ denote an
optimal {\em extreme point solution} to this LP. By reducing degree bounds $b(S)$, if needed, we assume that {\em
  $x$ satisfies all degree bounds at equality} (the degree bounds may
therefore be fractional-valued). Let $\alpha := 24$.

\begin{definition}
  An edge $e\in E$ is said to be {\em local} for $S\in\lam$ if $e$ has
  at least one end-point in $S$ but is neither in $E(C)$ nor in
  $\delta(C) \cap \delta(S)$ for any grandchild $C$ of $S$. Let
  $\local(S)$ denote the set of local edges for $S$.
 A node $S\in\lam$ is said to be {\em good} if $|\local(S)| \leq \alpha$.
\end{definition}

\begin{figure}[h]
\begin{center}
\includegraphics[scale=0.75]{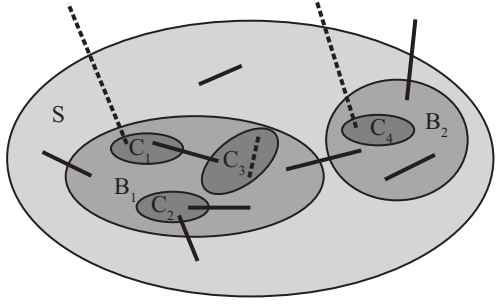}
\caption{Example of local edges.\label{fig:local-edges}}
\end{center}
\end{figure}

%\begin{floatingfigure}{7cm}
%  \includegraphics[scale=.6]{local-edges.eps}
%\end{floatingfigure}

Figure~\ref{fig:local-edges} shows a set $S$, its children $B_1$ and $B_2$, and grand-children $C_1, \ldots, C_4$;
edges in $\local(S)$ are drawn solid, non-local ones are shown dashed.

\begin{figure}[h]
\begin{center}
\includegraphics[scale=0.55]{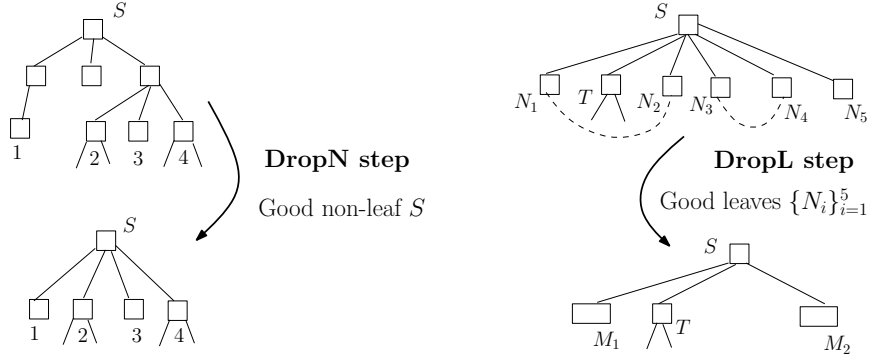}
\end{center}
%\vspace{-6mm}
\caption{\label{fig:mcst-iter} Examples of the degree constraint modifications DropN and DropL.}
%\vspace{-4mm}
\end{figure}

Initially, $E$ is the set of edges in the given graph, $F\leftarrow \emptyset$, $\lam$ is the original laminar family
of vertex sets for which there are degree bounds, and an arbitrary linear ordering is chosen on the children of each
node in $\mathcal{L}$. In a generic iteration $(F,E,\lam,b)$, the algorithm performs one of the following steps (see
also Figure~\ref{fig:mcst-iter}):
\begin{enumerate}
%\vspace{-2mm}
\item If $x_e=1$ for some edge $e\in E$ then $F\leftarrow F\cup\{e\}$,
  $E\leftarrow E\setminus\{e\}$, and set $b(S)\leftarrow b(S)-1$ for
  all $S\in\lam$ with $e\in\delta(S)$.
%\vspace{-2mm}
\item If $x_e=0$ for some edge $e\in E$ then $E\leftarrow
  E\setminus\{e\}$.
%\vspace{-2mm}
\item {\bf DropN:} Suppose there at least $|\lam|/4$ good non-leaf
  nodes in $\lam$.  Then either odd-levels or even-levels contain a
  set $\m\sse\lam$ of $|\lam|/8$ good non-leaf nodes.  Drop the degree
  bounds of all {\em children} of $\m$ and modify $\lam$
  accordingly. The ordering of siblings also extends naturally.
%\vspace{-2mm}
\item {\bf DropL:} Suppose there are more than $|\lam|/4$ good leaf
  nodes in $\lam$, denoted by $\n$. Then partition $\n$ into parts
  corresponding to siblings in $\lam$. For any part $\{N_1,\cdots,$
  $N_k\}\sse \n$ consisting of ordered (not necessarily contiguous)
  children of some node $S$:
%\vspace{-2mm}
 \begin{enumerate}
 \item Define $M_i=N_{2i-1}\cup N_{2i}$ for all $1\le i\le \lfloor
   k/2\rfloor$ (if $k$ is odd $N_k$ is not used).
 \item Modify $\lam$ by removing leaves $\{N_1,\cdots,N_k\}$ and
   adding new leaf-nodes $\{M_1,$ $\cdots,M_{\lfloor k/2\rfloor}\}$ as
   children of $S$ (if $k$ is odd $N_k$ is removed).  The children of
   $S$ in the new laminar family are ordered as follows: each node
   $M_i$ takes the position of either $N_{2i-1}$ or $N_{2i}$, and
   other children of $S$ are unaffected.
 \item Set the degree bound of each $M_i$ to
   $b(M_i)=b(N_{2i-1})+b(N_{2i})$.
\end{enumerate}
\end{enumerate}

Assuming that one of the above steps applies at each iteration, the algorithm terminates when $E=\emptyset$ and outputs
the final set $F$ as a solution. It is clear that the algorithm outputs a spanning tree of $G$. An inductive argument
(see e.g.~\cite{LNSS07}) can be used to show that the LP~\eqref{lp} is feasible at each each iteration and
$c(F)+z_{cur}\le z_o$ where $z_o$ is the original LP value, $z_{cur}$ is the current LP value, and $F$ is the chosen
edge-set at the current iteration. Thus the cost of the final solution is at most the initial LP optimum $z_o$. Next we
show that one of the four iterative steps always applies.

%\vspace{-2mm}
\begin{lemma}
  In each iteration, one of the four steps above applies.
\end{lemma}

%\vspace{-3mm}
%\marginpar{move proof to appendix if needed}
\begin{pf}
  Let $x^*$ be the optimal basic solution of \eqref{lp}, and
  suppose that the first two steps do not apply. Hence, we
  have $0 < x^*_e < 1$ for all $e \in E$. The fact that $x^*$
  is a basic solution together with a standard uncrossing
  argument (e.g., see \cite{J01})
  implies that $x^*$ is uniquely defined by
$$  x(E(U))  = |U| - |F(U)| - 1 \quad \forall\, U \in {\cal S}, \quad \mbox{ and }  \quad   x(\delta_E(S)) = b(S), \quad \forall \,S \in \lam',$$
%  \begin{align*}
%    x(E(U)) & = |U| - |F(U)| - 1 & \forall U \in {\cal S} \\
%    x(\delta_E(S)) & = b(S)  & \forall S \in \lam',
%  \end{align*}
  where ${\cal S}$ is a laminar subset of the tight spanning tree
  constraints, and $\lam'$ is a subset of tight degree constraints,
  and where     $|E|=|\mathcal S|+|\lam'|$.
%  \begin{equation}\label{eq:supsz}
%    |E|=|\mathcal S|+|\lam'|.
%  \end{equation}

  A simple counting argument (see,  e.g.,~\cite{SL07}) shows that there are at least $2$ edges induced
  on each $S\in{\cal S}$ that are not induced on any of its children; so
  $2|\mathcal S|\leq |E|$.
%Together with \eqref{eq:supsz} this implies  that
Thus we obtain $    |E| \leq 2 |\lam'| \leq 2 |\lam|$.

%  \begin{equation}\label{eq:esz}
%    |E| \leq 2 |\lam'| \leq 2 |\lam|.
%  \end{equation}
  From the definition of local edges, we get
  that any edge $e=(u,v)$ is local to at most the following six sets:
  the smallest set $S_1\in\lam$ containing $u$, the smallest set
  $S_2\in\lam$ containing $v$, the parents $P_1$ and $P_2$ of $S_1$
  and $S_2$ resp., the least-common-ancestor $L$ of $P_1$ and $P_2$,
  and the parent of $L$. Thus $\sum_{S\in\lam} |\local(S)| \leq
  6|E|$. From the above, we conclude that $\sum_{S\in\lam}
  |\local(S)| \leq 12|\lam|$. Thus at least $|\lam|/2$ sets $S\in\lam$
  must have $|\local(S)|\leq \alpha = 24$, i.e., must be good.  Now
  either at least $|\lam|/4$ of them must be non-leaves or at least
  $|\lam|/4$ of them must be leaves. In the first case, step 3 holds
  and in the second case, step 4 holds.
\end{pf}

It remains to bound the violation in the degree constraints, which turns out to be rather challenging. We note that
this is unlike usual applications of iterative rounding/relaxation, where the harder part is in showing that one of the
iterative steps applies.

It is clear that the algorithm reduces the size of $\lam$ by at least $|\lam|/8$ in each DropN or DropL iteration.
Since the initial number of degree constraints is at most $2n-1$, we get the following lemma.

%\vspace{-2mm}
\begin{lemma}\label{lem:mcst-iter-num}
The number of drop iterations (DropN and DropL) is $T := O(\log n)$.
\end{lemma}

%\vspace{-7mm}

\noindent {\bf Performance guarantee for degree constraints.}  We begin with some notation. The iterations of the
algorithm are broken into periods between successive drop iterations: there are exactly $T$ drop-iterations
(Lemma~\ref{lem:mcst-iter-num}). In what follows, the $t$-th drop iteration is called {\em round} $t$. The {\em time
$t$} refers to the instant just after round $t$; time $0$ refers to the start of the algorithm. At any time $t$,
consider the following parameters.
\begin{itemize}
\item $\lam_t$ denotes the laminar family of degree constraints.
\item $E_t$ denotes the undecided edge set, i.e., support of the
  current LP optimal solution.
\item For any set $\bs$ of {\em consecutive siblings} in $\lam_t$,
  $\bnd(\bs,t)=\sum_{N\in\bs} b(N)$ equals the sum of the residual
  degree bounds on nodes of $\bs$.
\item For any set $\bs$ of {\em consecutive siblings} in $\lam_t$,
  $\inc(\bs,t)$ equals the number of edges from
  $\delta_{E_t}(\cup_{N\in\bs}N)$ included in the final solution.
\end{itemize}

Recall that $b$ denotes the {\em residual} degree bounds at any point in the algorithm. The following lemma is the main
ingredient in bounding the degree violation.

\begin{lemma}
  \label{l:key} For any set $\bs$ of consecutive siblings in $\lam_t$
  (at any time $t$), $\inc(\bs,t)\le \bnd(\bs,t)+4\alpha\cdot (T-t)$.
\end{lemma}

Observe that this implies the desired bound on each original degree constraint $S$: using $t=0$ and $\bs=\{S\}$, the
violation is bounded by an additive $4\alpha \cdot T$ term.\\

%\vspace{-2mm}
\begin{pf} The proof of this lemma is by induction on $T-t$. The base
  case $t=T$ is trivial since the only iterations after this
  correspond to including 1-edges: hence there is no violation in {\em
    any} degree bound, i.e. $\inc(\{N\},T)\le b(N)$ for all $N\in
  \lam_T$. Hence for {\em any} $\bs\sse\lam$, $\inc(\bs,T)\le \sum_{N\in\bs}
  \inc(\{N\},T)\le \sum_{N\in\bs} b(N)=\bnd(\bs,T)$.

Now suppose $t<T$, and assume the lemma for $t+1$. Fix a consecutive $\bs\sse \lam_t$.
We consider different cases depending on what kind of drop occurs in round $t+1$.\\

%\vspace{-2mm}
  \noindent{\bf DropN round.} Here either all nodes in $\bs$ get
  dropped or none gets dropped.

  Case 1: {\em None of $\bs$ is dropped.} Then observe that $\bs$ is consecutive
  in $\lam_{t+1}$ as well; so the inductive hypothesis
  implies $\inc(\bs,t+1)\le \bnd(\bs,t+1)+4\alpha\cdot (T-t-1)$. Since
  the only iterations between round $t$ and round $t+1$ involve
  edge-fixing, we have $\inc(\bs,t)\le \bnd(\bs,t) -
  \bnd(\bs,t+1)+\inc(\bs,t+1)\le \bnd(\bs,t)+4\alpha\cdot (T-t-1)\le \bnd(\bs,t)+4\alpha\cdot (T-t)$.

  Case 2: {\em All of $\bs$ is dropped.} Let $\cs$ denote the set of
  all children (in $\lam_t$) of nodes in $\bs$. Note that $\cs$
  consists of consecutive siblings in $\lam_{t+1}$, and inductively
  $\inc(\cs,t+1)\le \bnd(\cs,t+1)+4\alpha\cdot (T-t-1)$. Let
  $S\in\lam_t$ denote the parent of the $\bs$-nodes; so $\cs$ are
  grand-children of $S$ in $\lam_t$.  Let $x$ denote the optimal LP
  solution {\em just before} round $t+1$ (when the degree bounds are
  still given by $\lam_t$), and $H=E_{t+1}$ the support edges of
  $x$. At that point, we have $b(N)=x(\delta(N))$ for all $N\in\bs\cup
  \cs$. Also let $\bnd'(\bs,t+1):= \sum_{N\in\bs} b(N)$ be the sum of
  bounds on $\bs$-nodes just before round $t+1$. Since $S$ is a good
  node in round $t+1$, $| \bnd'(\bs,t+1)- \bnd(\cs,t+1)| =
  |\sum_{N\in\bs} b(N) - \sum_{M\in\cs} b(M)|=|\sum_{N\in\bs}
  x(\delta(N))-\sum_{M\in\cs} x(\delta(M))|\le 2\alpha$. The last
  inequality follows since $S$ is good; the factor of $2$ appears
  since some edges, e.g., the edges between two children or two
  grandchildren of $S$, may get counted twice. Note also that the
  symmetric difference of $\delta_{H}(\cup_{N\in\bs} N)$ and
  $\delta_{H}(\cup_{M\in\cs} M)$ is contained in $\local(S)$. Thus
  $\delta_{H}(\cup_{N\in\bs} N)$ and $\delta_{H}(\cup_{M\in\cs} M)$
  differ in at most $\alpha$ edges.

  Again since all iterations between time $t$ and $t+1$ are
  edge-fixing:
%\vspace{-3mm}
 \begin{eqnarray*}
   \inc(\bs,t)&\le &\bnd(\bs,t) - \bnd'(\bs,t+1) + |\delta_{H}(\cup_{N\in\bs} N)\setminus \delta_{H}(\cup_{M\in\cs} M)| \\
   && \quad + \inc(\cs,t+1) \\
   &\le& \bnd(\bs,t) - \bnd'(\bs,t+1) + \alpha + \inc(\cs,t+1) \\
   &\le &\bnd(\bs,t) - \bnd'(\bs,t+1) + \alpha +\bnd(\cs,t+1)+4\alpha\cdot (T-t-1)\\
   &\le &\bnd(\bs,t) - \bnd'(\bs,t+1) + \alpha + \bnd'(\bs,t+1)+ 2\alpha + 4\alpha\cdot (T-t-1)\\
   &\le &\bnd(\bs,t) + 4\alpha\cdot (T-t)
 \end{eqnarray*}
%\vspace{-2mm}

 The first inequality above follows from simple counting; the second
 follows since $\delta_{H}(\cup_{N\in\bs} N)$ and
 $\delta_{H}(\cup_{M\in\cs} M)$ differ in at most $\alpha$ edges; the
 third is the induction hypothesis, and the fourth is
 $\bnd(\cs,t+1)\le \bnd'(\bs,t+1)+2\alpha$ (as shown above).

%\vspace{2mm}
 \noindent {\bf DropL round.} In this case, let $S$ be the parent of
 $\bs$-nodes in $\lam_t$, and $\n=\{N_1,\cdots,N_p\}$ be all the
 ordered children of $S$, of which $\bs$ is a subsequence (since it is
 consecutive). Suppose indices $1\le \pi(1)<\pi(2)<\cdots<\pi(k)\le p$
 correspond to good leaf-nodes in $\n$. Then for each $1\le i\le
 \lfloor k/2\rfloor$, nodes $N_{\pi(2i-1)}$ and $N_{\pi(2i)}$ are
 merged in this round. Let $\{\pi(i)\mid e\le i\le f\}$ (possibly
 empty) denote the indices of good leaf-nodes in $\bs$. Then it is
 clear that the only nodes of $\bs$ that may be merged with nodes
 outside $\bs$ are $N_{\pi(e)}$ and $N_{\pi(f)}$; all other
 $\bs$-nodes are either not merged or merged with another
 $\bs$-node. Let $\cs$ be the inclusion-wise minimal set of {\em
   children of $S$ in $\lam_{t+1}$} s.t.
\begin{itemize}

%\vspace{-2mm}
\item $\cs$ is consecutive in $\lam_{t+1}$,

%\vspace{-2mm}
\item $\cs$ contains all nodes of $\bs\setminus \{N_{\pi(i)}\}_{i=1}^k$, and
  % Rohit: I have reverted third bullet to version 8 (check!)

%\vspace{-2mm}
\item $\cs$ contains all new leaf nodes resulting from merging {\em two good leaf nodes} of $\bs$.
% old version commented below
% \item $\cs$ contains all new leaf nodes resulting from
%   $\{N_{\pi(i)}\mid e+1\le i\le f-1\}$ (i.e. those produced by
%   merging {\em two good leaf nodes} of $\bs$).
\end{itemize}
%\vspace{-2mm}

Note that $\cup_{M\in\cs} M$ consists of some subset of $\bs$ and at most two good leaf-nodes in $\n\setminus\bs$.
These two extra nodes (if any) are those merged with the good leaf-nodes $N_{\pi(e)}$ and $N_{\pi(f)}$ of $\bs$.  Again
let $\bnd'(\bs,t+1) := \sum_{N\in\bs} b(N)$ denote the sum of bounds on $\bs$ just before drop round $t+1$, when degree
constraints are $\lam_t$.  Let $H=E_{t+1}$ be the undecided edges in round $t+1$. By the definition of bounds on merged
leaves, we have $\bnd(\cs,t+1)\le \bnd'(\bs,t+1) + 2\alpha$. The term $2\alpha$ is present due to the two extra good
leaf-nodes described above.

%\vspace{-2mm}
\begin{cl}\label{cl:leaf-drop}
We have $|\delta_{H}(\cup_{N\in\bs} N)\setminus \delta_{H}(\cup_{M\in\cs} M)| \leq 2\alpha$.
\end{cl}

%\vspace{-2mm}
\begin{pf}
  We say that $N\in \n$ is represented in $\cs$ if either $N\in \cs$
  or $N$ is contained in some node of $\cs$. Let $\mathcal{D}$ be set
  of nodes of $\bs$ that are {\em not} represented in $\cs$ and the
  nodes of $\n\setminus\bs$ that are represented in $\cs$. Observe
  that by definition of $\cs$, the set $\mathcal{D}\sse
  \{N_{\pi(e-1)},N_{\pi(e)},N_{\pi(f)},N_{\pi(f+1)}\}$; in fact it can
  be easily seen that $|\mathcal{D}|\le 2$. Moreover $\mathcal{D}$
  consists of only good leaf nodes. Thus, we have $|\cup_{L\in
    \mathcal{D}} \delta_H(L)| \leq 2\alpha$. Now note that the edges
  in $\delta_{H}(\cup_{N\in\bs} N)\setminus \delta_{H}(\cup_{M\in\cs}
  M)$ must be in $\cup_{L\in \mathcal{D}} \delta_H(L)$. This completes
  the proof. \end{pf}

%\vspace{-1mm}

As in the previous case, we have:
%\vspace{-3mm}
\begin{eqnarray*}
  \inc(\bs,t)&\le &\bnd(\bs,t) - \bnd'(\bs,t+1) + |\delta_{H}(\cup_{N\in\bs} N)\setminus \delta_{H}(\cup_{M\in\cs} M)| \\
  && \quad + \inc(\cs,t+1) \\
  &\le &\bnd(\bs,t) - \bnd'(\bs,t+1) + 2\alpha + \inc(\cs,t+1) \\
  &\le &\bnd(\bs,t) - \bnd'(\bs,t+1) + 2\alpha + \bnd(\cs,t+1)+ 4\alpha\cdot (T-t-1)\\
  &\le &\bnd(\bs,t) - \bnd'(\bs,t+1) + 2\alpha + \bnd'(\bs,t+1)+ 2\alpha + 4\alpha\cdot (T-t-1)\\
  &= &\bnd(\bs,t) + 4\alpha\cdot (T-t)
\end{eqnarray*}

%\vspace{-2mm}
The first inequality follows from simple counting; the second uses Claim~\ref{cl:leaf-drop}, the third is the induction
hypothesis (since $\cs$ is consecutive), and the fourth is $\bnd(\cs,t+1)\le \bnd'(\bs,t+1)+2\alpha$ (from above).

\noindent This completes the proof of the inductive step and hence Lemma~\ref{l:key}.
\end{pf}

\section{Hardness Results}
\label{sec:hard}

\ifthenelse{\boolean{full}}{
  In this section we prove Theorem~\ref{th:mcst-hard}; i.e. unless \np~has
  quasi-polynomial time algorithms, there is no polynomial time $O(\log^c m)$
  additive approximation for degree bounds for the minimum crossing spanning tree problem, where $c>0$
  is some universal constant. This result also holds in the absence of
  edge-costs. We note that this hardness result only holds for the
  general MCST problem, and not the laminar MCST addressed earlier.
}{
  We now prove Theorem \ref{th:mcst-hard}.
} The first step to proving this result is a hardness for the more general minimum crossing matroid basis problem:
given a matroid $\m$ on a ground set $V$ of elements, a cost function $c:V\rightarrow\mathbb{R}_+$, and degree bounds
specified by pairs $\{(E_i,b_i)\}_{i=1}^m$ (where each $E_i\sse V$ and $b_i\in \mathbb{N}$), find a minimum cost basis
$I$ in $\m$ such that $|I\cap E_i|\le b_i$ for all $i\in [m]$.

\begin{theorem}
  \label{th:mat-hard} Unless \np~has quasi-polynomial time algorithms,
  the unweighted minimum crossing matroid basis problem admits no
  polynomial time $O(\log^c m)$ additive approximation for the degree bounds
  for some fixed constant $c>0$.
\end{theorem}
\begin{pf}
  We reduce from the label cover problem~\cite{A+93}. The input is a
  graph $G=(U,E)$ where the vertex set $U$ is partitioned into pieces
  $U_1,\cdots,U_n$ each having size $q$, and all edges in $E$ are
  between distinct pieces.  We say that there is a {\em superedge}
  between $U_i$ and $U_j$ if there is an edge connecting some vertex
  in $U_i$ to some vertex in $U_j$. Let $t$ denote the total number of
  superedges; i.e.,

  $$t=\left|\left\{(i,j)\in {[n]\choose 2}: \mbox{ there is an edge in $E$
        between $U_i$ and $U_j$}\right\}\right|$$

  The goal is to pick one vertex from each part $\{U_i\}_{i=1}^n$ so
  as to maximize the number of induced edges. This is called the value
  of the label cover instance and is at most $t$.

  It is well known that there exists a universal constant $\gamma>1$
  such that for every $k\in \mathbb{N}$, there is a reduction from any
  instance of SAT (having size $N$) to a label cover instance $\langle
  G=(U,E),q,t\rangle$ such that:
\begin{itemize}
\item If the SAT instance is satisfiable, the label cover instance
  has optimal value $t$.

\item If the SAT instance is not satisfiable, the label cover instance
  has optimal value $< t/\gamma^k$.

\item $|G|=N^{O(k)}$, $q=2^k$, $|E|\le t^2$, and the reduction runs in time $N^{O(k)}$.
\end{itemize}

We consider a uniform matroid $\m$ with rank $t$ on ground set $E$ (recall that any subset of $t$ edges is a basis in a
uniform matroid). We now construct a crossing matroid basis instance \is on $\m$. There is a set of degree bounds
corresponding to each $i\in [n]$: for every collection $C$ of edges incident to vertices in $U_i$ such that no two
edges in $C$ are incident to the same vertex in $U_i$, there is a degree bound in \is requiring {\em at most one}
element to be chosen from $C$. Note that the number of degree bounds $m$ is at most $|E|^q\le \smash{N^{O(k \,2^k)}}$.
The following claim links the SAT and crossing matroid instances. \ifthenelse{\boolean{full}}{}{Its proof is deferred
to the full
  version of this paper.}

\begin{cl}
  \noindent{\em [\underline{Yes instance}]} If the SAT instance is satisfiable, there
  is a basis (i.e. subset $B\sse E$ with $|B|=t$) satisfying all degree
  bounds. \\ \noindent {\em [\underline{No instance}]} If the SAT instance is unsatisfiable,
  every subset $B'\sse E$ with $|B'|\ge t/2$ violates some degree
  bound by an additive $\rho=\gamma^{k/2}/\sqrt{2}$.
\end{cl}
\ifthenelse{\boolean{full}}{\begin{pf}
  Observe that if the original SAT instance is satisfiable, then the
  matroid $\m$ contains a basis obeying all the degree bounds: namely
  the $t$ edges $T^*\sse E$ covered in the optimal solution to the
  label cover instance. This is because if we consider any $U_i$, then
  all the $T^*$-edges having a vertex in $U_i$ as their endpoint, have
  the same endpoint.  Thus, for any degree bound corresponding to
  collection $C$ (as defined above), at most one $T^*$-edge can lie in
  $C$.

  Now consider the case that the SAT instance is unsatisfiable. Let
  $B'\sse E$ be any subset with $|B'|\ge t/2$. We claim that $B'$
  contains at least $\rho=\gamma^{k/2}/\sqrt{2}$ edges from some
  degree-constrained set of edges. Suppose (for a contradiction) that
  $|B'\cap C|< \rho$ for each degree constraint $C$. This means that
  each part $\{U_i\}_{i=1}^n$ contains fewer than $\rho$ vertices that
  are incident to edges $B$. For each part $i\in [n]$, let $W_i\sse
  U_i$ denote the vertices incident to edges of $B$; note that $|W_i|<
  \rho$. Consider the label cover solution obtained as follows.  For
  each $i\in[n]$, choose one vertex from $W_i$ independently and
  uniformly at random. Clearly, the expected number of edges in the
  resulting induced subgraph is at least $|B'|/\rho^2 \ge
  \frac{t}{2\rho^2} = t/\gamma^k$. This contradicts the fact that the
  value of label cover instance is strictly less than $t/\gamma^k$.
\end{pf}}{}

The steps described in the above reduction can be done in time polynomial in $m$ and $|G|$. Also, instead of randomly
choosing vertices from the sets $W_i$, we can use conditional expectations to derive a deterministic algorithm that
recovers at least $t/\rho^2$ edges. Setting $k=\Theta(\log\log N)$ (recall that $N$ is the size of the original SAT
instance), we obtain an instance of bounded-degree matroid basis of size $\max\{m,|G|\} = N^{\log^a N}$ and
$\rho=\log^b N$, where $a,b>0$ are constants. Note that $\log m= \log^{a+1}N$, which implies $\rho=\log^c m$ for
$c=\frac{b}{a+1}>0$, a constant. Thus it follows that for this constant $c>0$ the bounded-degree matroid basis problem
has no polynomial time $O(\log^c m)$ {\em
  additive approximation} for the degree bounds, unless \np~has quasi-polynomial time
algorithms.\end{pf}

We now prove Theorem~\ref{th:mcst-hard}.

\begin{pf}[Proof of Theorem \ref{th:mcst-hard}]
  We show how the bases of a uniform matroid can be represented in a
  suitable instance of the crossing spanning tree problem. Let the
  uniform matroid from Theorem~\ref{th:mat-hard} consist of $e$
  elements and have rank $t\le e$; recall that $t\ge \sqrt{e}$ and
  clearly $m\le 2^e$. We construct a graph as in
  Figure~\ref{fig:xsp-hard}, with vertices $v_1,\cdots,v_e$
  corresponding to elements in the uniform matroid. Each vertex $v_i$
  is connected to the root $r$ by two vertex-disjoint paths: $\langle
  v_i,u_i,r\rangle$ and $\langle v_i,w_i,r\rangle$. There are no costs
  in this instance.  Corresponding to each degree bound (in the
  uniform matroid) of $b(C)$ on a subset $C\sse [e]$, there is a
  constraint to pick at most $|C|+b(C)$ edges from $\delta(\{u_i\mid
  i\in C\})$. Additionally, there is a {\em special degree bound} of
  $2e-t$ on the edge-set $E'=\bigcup_{i=1}^e \delta(w_i)$; this
  corresponds to picking a basis in the uniform matroid.

\ifthenelse{\boolean{full}}{
  \begin{figure}[h!]
    \begin{center}
      \includegraphics[scale=0.6]{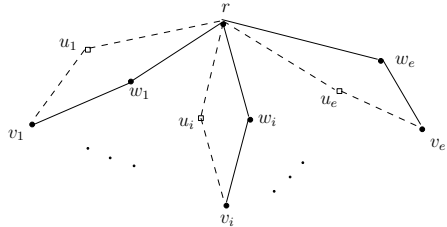}
    \end{center}
    \caption{\label{fig:xsp-hard} The crossing spanning tree instance used in the reduction.}
  \end{figure}
}{
  \piccaptioninside
  \piccaption{\label{fig:xsp-hard} The crossing spanning tree instance used in the reduction.}
  \parpic(6cm,3.5cm)[fr]{
    \includegraphics[scale=.5]{xsp-hard.eps}}
}

Observe that for each $i \in [e]$, any spanning tree must choose exactly three edges amongst $\{(r,u_i),(u_i,v_i),
\\(r,w_i),(w_i,v_i)\}$, in fact any three edges suffice. Hence every
spanning tree $T$ in this graph corresponds to a subset $X\sse[e]$ such that: (I) $T$ contains both edges in
$\delta(u_i)$ and one edge from $\delta(w_i)$, for each $i\in X$, and (II) $T$ contains both edges in $\delta(w_i)$ and
one edge from $\delta(u_i)$ for each $i\in [e]\setminus X$.

From Theorem~\ref{th:mat-hard}, for the crossing matroid problem, we obtain the two cases:

\smallskip
\noindent\underline{\em Yes instance.} There is a basis $B^*$ (i.e. $B^*\sse [e]$,
  $|B^*|=t$) satisfying all degree bounds. Consider the spanning tree
  $$ T^*=\{(r,u_i), (u_i,v_i), (r,w_i)\mid i\in B^*\} \bigcup \{(r,w_i),
  (u_i,w_i), (r,u_i)\mid i\in [e]\setminus B^*\}.$$
  Since $B^*$
  satisfies its degree-bounds, $T^*$ satisfies all degree bounds
  derived from the crossing matroid instance. For the special degree
  bound on $E'$, note that $|T^*\cap E'|=2e-|B^*|=2e-t$; so this is
  also satisfied. Thus there is a spanning tree satisfying all the
  degree bounds.

\smallskip
\noindent\underline{\em No instance.} Every subset $B'\sse [e]$ with $|B'|\ge t/2$ (i.e. near basis) violates some
degree bound by an additive $\rho=\Omega(\log^c m)$ term, where $c>0$ is a fixed constant. Consider any spanning tree
$T$ that corresponds to subset $X\sse [e]$ as described above.
\begin{enumerate}
 \item Suppose that $|X|\le t/2$; then we have $|T\cap E'| = 2e-|X|\ge 2e-t +\frac{t}2$, i.e. the special degree bound is
violated by $t/2\ge \Omega(\sqrt{e}) = \Omega(\log^{1/2} m)$.
 \item Now suppose that $|X|\ge t/2$. Then by the guarantee on the no-instance, $T$ violates some degree-bound derived
 from the crossing matroid instance by additive $\rho$.
\end{enumerate}
Thus in either case, every spanning tree violates some degree bound by additive $\rho=\Omega(\log^c m)$.

\smallskip
By Theorem~\ref{th:mat-hard}, it is hard to distinguish the above cases and we obtain the corresponding hardness result
for crossing spanning tree, as claimed in Theorem~\ref{th:mcst-hard}.
\end{pf}

%%%%%%%%%%%%%%%%%%%%%%%%%%%%%%%%%%%%%%%%%%%%%%%%%%%%%%%%%%%%%%%%%%%%%%%%%%%%%%%%%%%%%%%%%%
\ifthenelse{\boolean{full}}{  %%%%%%%%% Full Version Only

\subsection{Hardness for Robust $k$-median}
\label{sec:kmed-hard}

Another interesting consequence of Theorem~\ref{th:mat-hard} is for the robust $k$-median problem~\cite{AGGN08}. Here
we are given a metric $(V,d)$, $m$ client-sets $\{S_i\sse V\}_{i=1}^m$, and bound $k$; the goal is to find a set $F\sse
V$ of $k$ facilities such that the worst-case connection cost (over all client-sets) is minimized, i.e.
$$\min_{F\sse V, |F|=k} \max_{i=1}^m \sum_{v\in S_i} d(v,F).$$
Above $d(v,F)$ denotes the shortest distance from $v$ to any vertex in $F$. Anthony et al.~\cite{AGGN08} gave an
$O(\log m +\log k)$-approximation algorithm for robust $k$-median, and showed that it is hard to approximate better
than factor two. At first sight this problem may seem unrelated to crossing matroid basis. However using
Theorem~\ref{th:mat-hard}, we obtain the poly-logarithmic hardness result stated in Corollary~\ref{cor:k-med-hard}.

\begin{pf}
Recall that in a uniform metric, the distance between every pair of vertices is one. In this case the robust $k$-median
problem can be rephrased as:
$$\min_{F\sse V, |F|=k} \, \max_{i=1}^m \,\,|S_i\setminus F|,\quad \mbox{ where $\{S_i\sse V\}_{i=1}^m$ are the client-sets}.$$

The hard instances of crossing matroid basis in Theorem~\ref{th:mat-hard} are in fact for uniform matroids where every
degree upper-bound equals {\em one}. i.e. there is a ground-set $V$, degree bounds given by $\{E_i\sse V\}_{i=1}^m$,
and rank $t$; the goal is to find (if possible) a subset $I\sse V$ with $|I|=t$ such that $|I\bigcap E_i|\le 1$ for all
$i\in[m]$. Theorem~\ref{th:mat-hard} showed that it is hard to distinguish the following cases: (Yes-case) there is
some $I\sse V$ with $|I|=t$ and $\max_{i\in[m]} |I\cap E_i| \le 1$; and (No-case) for every $I\sse V$ with $|I|=t$,
$\max_{i\in[m]} |I\cap E_i| \ge \rho :=\Omega(\log^c m)$.

These hard instances naturally correspond to the robust $k$-median problem on uniform metric $V$, client-sets
$\{E_i\sse V\}_{i=1}^m$, and bound $k=|V|-t$. It is clear that the robust $k$-median objective is at most one in the
Yes-case, and at least $\rho$ in the No-case. Thus we obtain a multiplicative $\rho$ hardness of approximation for
robust $k$-median on uniform metrics. This proves Corollary~\ref{cor:k-med-hard}.
\end{pf}

\subsection{Integrality Gap for general MCST}
\label{sec:mcst-gap}

We now present the $b+\Omega(\sqrt{n})$ integrality gap instance for minimum crossing spanning tree.  While such gaps
instances are easy to obtain if one allows $m$ to be super-polynomially large (for example, by setting a degree bound
for each subset of edges), the nice property of the example here is that $m$ is quite small, in fact $m=O(n)$. This
result is due to Mohit Singh~\cite{S08}, we thank him for letting us present the example here.

The graph is the same as the one used for the hardness result. The vertex-set is $\{r\}\bigcup \{v_i,u_i,w_i\}_{i=1}^e$
%$\{r\}\bigcup \{v_i\}_{i=1}^e\bigcup \{u_i\}_{i=1}^e\bigcup \{w_i\}_{i=1}^e$;
so $n=3e+1$. The edges are $\{(r,u_i)\mid i\in [e]\}\cup \{(v_i,u_i)\mid i\in [e]\}$ and $\{(r,w_i)\mid i\in [e]\}\cup
\{(v_i,w_i)\mid i\in [e]\}$. See also Figure~\ref{fig:xsp-hard}. There are no costs in this instance.

The `degree bounds' for the MCST instance are derived from the lower bound for the {\em discrepancy
problem}~\cite{C00}. From discrepancy theory there exists a collection $\{S_j\sse [e] \}_{j=1}^e$ of subsets such that,
$$\max_{j=1}^e \big| |X\cap S_j| - |\overline{X}\cap S_j|\big| \ge \rho, \quad \mbox{for every }X\sse [e].$$
Above $\overline{X}=[e]\setminus X$ as usual, and $\rho=\Omega(\sqrt{e})=\Omega(\sqrt{n})$. In other words, for every
way of partitioning $[e]$, there is some set $S_j$ such that the partition induced on $S_j$ has a large imbalance.
There are $m=2e$ degree bounds, defined as follows. For each $j\in[e]$ there is a bound of $|S_j|+\lceil |S_j|/2
\rceil$ on each of the edge-sets $U_j =\cup_{i\in S_j} \delta(u_i)= \{(r,u_i),(u_i,v_i)\}_{i\in S_j}$,  and $W_j
=\cup_{i\in S_j} \delta(w_i)= \{(r,w_i),(w_i,v_i)\}_{i\in S_j}$.

Consider the fractional solution to the natural LP relaxation that sets each edge to value $3/4$. It is easily seen
that it is indeed a fractional spanning tree and satisfies all the degree bounds.

On the other hand, we claim that any integer solution must violate some degree bound by additive $\frac{\rho}2-1$. Note
that every spanning tree $T$ in this graph corresponds to a subset $X\sse[e]$ such that: (I) $T$ contains both edges in
$\delta(u_i)$ and one edge from $\delta(w_i)$, for each $i\in X$, and (II) $T$ contains both edges in $\delta(w_i)$ and
one edge from $\delta(u_i)$ for each $i\in \overline{X}$. The number of edges used by tree $T$ in the degree-bounds
(for each $j\in[e]$) are:
\begin{itemize}
\item  $|T\cap U_j| = 2\, |X\cap S_j| + |\overline{X}\cap S_j| = |S_j| + |X\cap S_j|$, and
\item $|T\cap W_j| =  |X\cap S_j| + 2\,|\overline{X}\cap S_j| = |S_j| + |\overline{X}\cap S_j|$.
\end{itemize}

From the discrepancy instance, it follows that $\max_{j=1}^e \big| |X\cap S_j| - |\overline{X}\cap S_j|\big| \ge \rho$;
let $k$ be the index achieving this maximum. Then we have:
$$\max\{ |T\cap U_k|,\, |T\cap W_k| \} = |S_k| + \max\{|X\cap S_k|,\, |\overline{X}\cap S_k| \} \ge |S_k|+\frac{|S_k|}2 +
\frac{\rho}2.$$ Thus the degree-bound for either $U_k$ or $W_k$ is violated by additive $\frac{\rho}2-1$.

\section{Minimum Crossing Contra-Polymatroid Intersection}
\label{sec:mcpi}

In this section we consider the {\em  crossing contra-polymatroid intersection problem} (see
Definition~\ref{def.BoundedPolymatroidIntersection}) and prove Theorem~\ref{t.matroidintersection}. The algorithm
(given as Algorithm~\ref{alg2}) for this problem is based on iteratively relaxing the following natural LP relaxation.
\begin{align*}
 \min\ &\sum_{e\in E'} c_e\cdot x_e\\
  &x(S\cap E')\ge  r_1(S) -|F\cap S| &\forall S\subseteq E\\
  &x(S\cap E')\ge r_2(S)-|F\cap S| &\forall S\subseteq E\\
  &x(E_i\cap E')\le b'_i &\forall i\in W\\
  &0\le x_e\le 1 &\forall e\in E'.
\end{align*}
At a generic iteration, $E'\sse E$ denotes the set of unfixed elements, $F\sse E$ the set of chosen elements (recall
that $E$ denotes the groundset of the instance), $W\sse I$ the set of remaining degree bounds, and $b'_i$ (for each
$i\in W$) the residual degree-bound in the $i^{th}$ constraint. Observe that this LP can indeed be solved in polynomial
time by the Ellipsoid algorithm: the separation oracle for the first two sets of constraints involve submodular
function minimization for the two functions $g_i(S) = x(S\cap E') + |S\cap F| - r_i(S)$ (with $i=1,2$). The resulting
fractional solution can then be converted to an extreme point solution of no larger cost, as described in
Jain~\cite{J01}.

\begin{algorithm}
\caption{Algorithm for minimum crossing contra-polymatroid intersection.\label{alg2}}
\begin{algorithmic}[1]
 \STATE Initially, set $E'=E$, $F=\emptyset$, $W=I$, $b'_i=b_i$, for all $i\in I$
 \WHILE{$E'\neq \emptyset$}
 \STATE Compute an optimal extreme point solution  $x^*$ of the LP$(E',F,W)$;
 \FORALL{$e\in E'$ with $x^*(e)=0$}
 \STATE $E'\leftarrow E'\setminus{\{e\}}$
 \ENDFOR
 \FORALL{$e\in E'$ with $x^*(e)\ge\frac{1}{2}$}
 \STATE $F\leftarrow F\cup\{e\}$;  $E'\leftarrow E'\setminus{\{e\}}$
 \STATE $b'_i\leftarrow b'_i - x^*(e)$, for all $i\in W$ with $e\in E_i$
 \ENDFOR \ignore{\item \tab FOR all $i\in I$
with $b_i-|F\cap E_i|\le 0$, $I\leftarrow I\setminus{\{i\}}$;}
 \FORALL{$i\in W$ with $|E_i\cap E'|\le \lceil 2b'_i\rceil +\Delta- 1$}
 \STATE $W\leftarrow W\setminus \{i\}$ \ENDFOR \ENDWHILE
 \STATE Return the incidence vector of $F$;
\end{algorithmic}
\end{algorithm}

Note that this algorithm rounds variables of value $x^*(e)\ge \frac{1}{2}$ to 1, and hence we loose a factor of two in
the cost and in the degree bounds. Theorem~\ref{t.matroidintersection} follows as a consequence if we can show that in
each iteration, either some variable can be rounded, or some constraint can be dropped.
%For this, we first prove:

\begin{lemma}\label{lem:mat-intersection}
If $x^*\in \R^E$ is an optimal extreme point solution to the above LP for crossing contra-polymatroid intersection,
with $0<x^*(e)<\frac{1}{2}$ for all $e\in E$, then there exists $i\in W$ such that
$$|E_i\cap E'|\le \lceil 2b'_i\rceil + \Delta -1$$
\end{lemma}
\begin{pf}
Let $\T'_i = \{ \chi(E'\cap S) | x^*(S\cap E') = r_1(S) - |S\cap F|,\,\, S\sse E\}$ for $i=1,2$ denote the tight sets
from the first two constraints of the LP. Let $\B' = \{\chi(E'\cap E_i) | x^*(E_i\cap E')=b'_i, \,\,i\in W\}$ denote
the tight degree constraints. Since $x^*$ is an extreme point solution (and $0<x^*<1$), there exist linearly
independent tight sets $\T_1\sse \T'_1$, $\T_2\sse \T'_2$ and $\B\sse \B'$ such that $|E'|=|\T_1| +|\T_2| +|\B|$.

\ignore{corresponding to the following constraints $\T_1\subseteq \{S\subseteq E\mid x^*(S\cap E') = r_1(S) - |S\cap
F|\}$, $\T_2\subseteq \{S\subseteq E\mid x^*(S\cap E')=r_2(S) - |S\cap F|\}$ and $\B\subseteq \{E_i\subseteq E\mid
x^*(E_i\cap E')=b'_i\}$ such that
$$|E|=|\T_1| +|\T_2| +|\B|.$$}

Since $x^*$ is modular and $r_i(S)-|S\cap F|$ (for $i=1,2$) are supermodular on $2^E$,
%the Boolean lattice $(2^E,\subseteq)$,
it can be assumed (again, using uncrossing arguments) that each of $(\T_1,\subseteq)$ and $(\T_2,\subseteq)$ forms a
chain\footnote{A family $(\lam,\sse)$ is a chain iff for every $X,Y\in\lam$, either $X\sse Y$ or $Y\sse X$.}. The
following claim goes back to a similiar result for spanning trees as stated in~\cite{BKN08}.
\begin{cl} \label{cl:mat-tok} For each $i=1,2$,  we have $|\T_i|\le x^*(E')$; additionally
if $|\T_i| = x^*(E')$ then $E'\in \T_i$.
\end{cl}

\begin{pf}
We prove the claim for $i=1$. Let $\T_1=\{S_1\subset \ldots \subset S_k\}$ where $S_k\sse E'$. Let $S_0=\emptyset$ and
consider an arbitrary pair of subsequent chain elements $S_i\subset S_{i+1}$, for any $i\in \{0,1,\ldots,k-1\}$. Since
$x^*_e>0$ for all $e\in E'$ it follows that $x^*(S_{i+1}\setminus{S_i})>0$. Hence, by the integrality of $r_1(S)-|S\cap
F|$ and tight constraints $S_i$ and $S_{i+1}$,
$$x^*(S_{i+1}\setminus{S_i})=x^*(S_{i+1})-x^*(S_i)=r_1(S_{i+1}) - |S_{i+1}\cap F| -r_1(S_i) + |S_i\cap F|\ge 1.$$
Summing over $i=0,\ldots,k-1$ we therefore obtain the inequality:
$$x^*(E')\ge x^*(S_k)=\sum_{i=0}^{k-1}x^*(S_{i+1}\setminus{S_i}) \ge k=|\T_1|,$$
with equality only if $E'=S_k$.
\end{pf}

We now proceed with the proof of Lemma~\ref{lem:mat-intersection}. Suppose (for a contradiction) that for all $i\in W$,
$|E_i\cap E'|\ge \lceil 2b'_i\rceil + \Delta$. For each $i\in W$, define $\s_i := \sum_{e\in E'\cap E_i} (1-2x^*_e) =
|E'\cap E_i|-2 x^*(E_i)$. Then we have $\s_i\ge |E'\cap E_i| - 2b'_i\ge |E'\cap E_i| - \lceil 2b'_i\rceil \ge \Delta$.
Hence $\sum_{i\in W} \s_i\ge \Delta\cdot |W|$.\ignore{We will show that $\sum_{i\in W} \s_i < \Delta\cdot |W|$ which is
the desired contradiction.}

For each $e\in E'$, let $r_e:= |\{i\in W: e\in E_i\}|\le \Delta$ the maximum element frequency. Note also that
$0<1-2x^*_e<1$ for each $e\in E'$. Now,
\begin{eqnarray*}
\sum_{i\in W} \s_i  & = & \sum_{e\in E'} r_e\cdot (1-2x^*_e)\le \Delta \cdot \sum_{e\in E'} (1-2x^*_e) \\
& = &  \Delta \cdot \left(|E'| - 2\cdot x^*(E')\right)\le \Delta \cdot \left(|E'| - |\T_1| - |\T_2|\right)
\end{eqnarray*}
The last inequality uses Claim~\ref{cl:mat-tok}. Note that equality holds above only if $E'\in \T_1\cap \T_2$ (by
Claim~\ref{cl:mat-tok}), which would contradict the linear independence of $\T_1$ and $\T_2$. Thus we have:
$$\sum_{i\in W} \s_i < \Delta \cdot \left(|E'| - |\T_1| - |\T_2|\right) = \Delta\cdot |\B|\le \Delta\cdot |W|.$$

\ignore{with equality only if $E'\in \T_1\cap \T_2$ (from Claim~\ref{cl:mat-tok}), $r_e=\Delta$ for all $e\in E'$, and
$\B=W$. We now claim that equality $\sum_{i\in W} \s_i = \Delta\cdot |W|$ is not possible. If this were the case,
$\chi(E')$ is a constraint in each of $\T_1$ and $\T_2$; and $\sum_{i\in \B} \chi(E_i) = \sum_{i\in W} \chi(E_i) =
\Delta\cdot \chi(E)$. However this contradicts the linear independence of constraints in $\T_1$ and $\B$.}

However this contradicts the assumption $|E'\cap E_i|\ge \lceil 2b'_i\rceil + \Delta$ for all $i\in W$.\end{pf}

\begin{pf}{\bf [Theorem~\ref{t.matroidintersection}]}
  Lemma~\ref{lem:mat-intersection} implies that an improvement is
  possible in each iteration of Algorithm~\ref{alg2}. Since we only
  round elements that the LP sets to value at least half, the cost
  guarantee is immediate. Consider any degree bound $i\in I$; let
  $b'_i$ denote its residual bound when it is dropped, and $F'$ (resp. $E'$) the
  set of chosen (resp. unfixed) elements  at that iteration. Again, rounding elements
  of fractional value at least half implies $|E_i\cap F'|\le \lfloor
  2b_i-2b'_i\rfloor = 2b_i - \lceil 2b'_i\rceil$. Furthermore, the
  number of $E_i$-elements in the support of the basic solution at the
  iteration (ie. $E'$) when constraint $i$ is dropped is at most $\lceil
  2b'_i\rceil + \Delta -1$. Thus the number of $E_i$-elements chosen
  in the final solution is  at most $|E_i\cap F'|+ |E_i\cap E'| \le 2b_i - \lceil 2b'_i\rceil + \lceil
  2b'_i\rceil + \Delta -1 = 2\cdot b_i+\Delta-1$.
\end{pf}

\medskip
\noindent {\bf Tight Example.} We note that the natural iterative relaxation steps (used above) are insufficient to
obtain a better approximation guarantee. Consider the special case of the crossing bipartite edge cover problem. The
instance consists of graph $G$ which is a $4n$-length cycle, with its edges partitioned into two perfect matchings
$E_1$ and $E_2$. There is a degree-bound of $n$ on each of $E_1$ and $E_2$; so $\Delta=1$. Consider the fractional
solution to the LP-relaxation that assigns value of $\frac12$ to all edges. It is indeed a fractional edge-cover since
each vertex is covered to extent one. The degree-bounds are clearly satisfied. It is also an extreme point: note that
this is the unique fractional solution minimizing the all-ones cost vector. For this extreme point solution, the
largest edge-value is $\frac12$, and the support-size (i.e. $2n$) of its degree-constraints is twice their bound (i.e.
$n$). Thus the iterative relaxation must either pick a half-edge or drop a degree-constraint that is potentially
violated by factor two.

\section{Minimum Crossing Lattice Polyhedra}
\label{sec:mclp}

Before formally defining the lattice polyhedra problem, we need to introduce some terminology. We use notation similar
to~\cite{Fra99}. Let $(\F,\le)$ be a partially ordered set with $\F\ne \emptyset$. \ignore{Two elements $A,B\in\F$
  are said to be {\em comparable} if either $A\le B$ or $B\le A$; they
  are {\em non-comparable} otherwise. A subset $\lam\sse \f$ is called
  a {\em chain} if $\lam$ contains no pair of non-comparable
  elements.} We consider a {\em lattice} $(\F,\le)$, where there are
two commutative binary operations, {\em meet} $\wedge$ and {\em join} $\vee$, that are defined on {\em all} pairs
$A,B\in\f$, such that:
$$A\wedge B~ \le ~A,B~\le ~A\vee B$$
Note that our definition is more general than the usual definition of a lattice, since the join $A\vee B$ is not
required to be the least common upper bound of $A$ and $B$. A function $r:\f\rightarrow \mathbb{Z}_+$ is said to be
{\em supermodular} on $(\F,\le, \wedge, \vee)$ iff:
$$r(A) + r(B) \le r(A\wedge B) + r(A\vee B),\quad \mbox{for all }A,B\in\f$$
Given a supermodular function $r:\f\rightarrow \mathbb{Z}_+$, a ground set $E$, a cost function $c:E\rightarrow
\mathbb{R}_+$, and a set-valued function $\rho:\f\rightarrow 2^E$ satisfying:
\begin{enumerate}
 \item {\bf Consecutive property:} If $A\le B\le C$ then $\rho(A)\cap \rho(C)\sse \rho(B)$,
 \item {\bf Submodularity:} For all $A,B\in\f$, $\rho(A\vee B)\cup \rho(A\wedge B)\sse \rho(A)\cup \rho(B)$,
% \item If $\rho(A)\cap \rho(B)\ne \emptyset$ then $A,B$ are intersecting or comparable.
\end{enumerate}
the {\em lattice polyhedron problem} is defined as the following integer program:
$$\min\left\{c^T\cdot x\mid \sum_{e\in \rho(S)} x_e\ge r(S),~\forall S\in\f;\quad x\in\{0,1\}^E\right\}.$$

\begin{definition}[Minimum crossing lattice polyhedron] \label{def:xlattp}
Given a lattice polyhedron $\langle E, (\F,\le), r,\rho,c\rangle$ as above, and lower/upper bounds $\{a_i\}_{i\in I}$
and $\{b_i\}_{i\in I}$ on a collection $\{E_i\subseteq E\}_{i\in I}$, the goal is to minimize:
$$\left\{c^T\cdot x\mid \sum_{e\in \rho(S)} x_e\ge r(S),~\forall S\in\f;\quad a_i\le x(E_i)\le b_i,~\forall i\in I;\quad x\in\{0,1\}^E\right\}.$$
\end{definition}

We already mentioned in the introduction that
%crossing matroid intersection as well as
several discrete optimization problems fit into the lattice polyhedron model (see e.g.~\cite{Sch03}).

For example, in the {\em contra-polymatroid intersection} problem with two supermodular rank functions $r_1, r_2:2^E\to
\R$, the lattice $(\F,\le)$ consists of two copies $S'$ and $S''$ for each subset $S\subseteq E$, with partial order:
$$S'\le T'' \quad \mbox{ and } \quad (S\subseteq T \implies S'\le T',~S''\ge T'');\qquad \forall~ S,T\subseteq 2^E.$$
This is easily seen to satisfy the consecutivity and submodularity properties. The rank function $r$ for the lattice
polyhedron has $r(S')=r_1(S)$ and $r(S'')=r_2(S)$, for all $S\sse E$.
%Certainly, $r$ is supermodular.)
%\emph{Example 1: Crossing matroid intersection.} Let $r_1,r_2:2^E\to \R$ be two supermodular rank functions. Then the
%crossing matroid intersection polyhedron can be written as
%$$\{x\in [0,1]^E\mid x(S)\ge r(S) \quad \forall S\in \F, \quad  x(E_i)\le b_i \quad \forall i\in I\}$$
%where
%$\F$ contains two copies $S'$ and $S''$ for each subset $S\subseteq E$
%and $r\in \R^{\F}$ has values $r(S')=r_1(S)$ and $r(S'')=r_2(S)$.

In the {\em planar min-cut} problem, recall that $\F$ consists of all $s-t$ paths in the given $s,t$-planar graph $G$.
The partial order sets for any pair of $s-t$ paths $P,Q$,
$$P\le Q \quad\Longleftrightarrow\quad P \mbox{ ``below'' }Q \mbox{ in the planar representation}.$$
The induced lattice turns out to be consecutive and submodular. The rank function is the all-ones function.
%(Slightly more formally, we could also write $P\le Q$ iff $Q$ is the \emph{uppermost} path in the subgraph containing only the edges from $P$ and $Q$.
%Here, the submodularity of the induced lattice follows immediatly from the order-definition, while the planarity is crucial for the consecutivity.
For more details on the relation between planar min cut and lattice polyhedra, the reader is referred to~\cite{FKP08}.

%\vspace{-3mm}
\subsection{Integrality gap for general crossing lattice polyhedra}
\label{sec:lph-gap}

We first show that there is a bad integrality gap for crossing lattice polyhedra. Consider the planar min-cut instance
on graph $G=(V,E)$ in Figure~\ref{fig:latt-gap} with vertices $s,t\in V$ as shown. Define edge-sets
$E_i:=\{(v_{i-1},u_{i,j})\}_{j=1}^k \bigcup \{(v_{i},u_{i,j})\}_{j=1}^k$ for each $i\in [k]$; here we set $v_0=s$ and
$v_k=t$. There are only degree upper-bounds in this instance, namely bound of one on each $\{E_i\}_{i=1}^k$. Note also
that $\Delta=1$ in this instance, and size of the ground-set $n=|E|=\Theta(k^2)$.

\begin{figure}[h!]
\begin{center}
\includegraphics[scale=0.7]{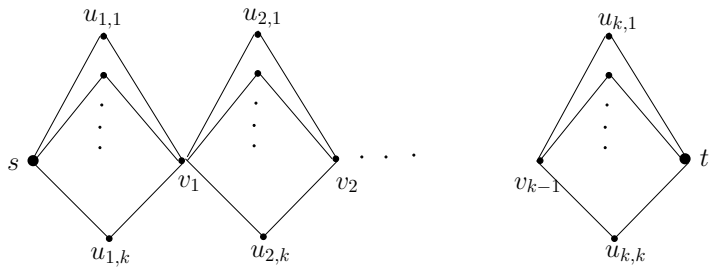}
\end{center}
\caption{\label{fig:latt-gap} The integrality gap instance for crossing planar min-cut.}
\end{figure}

Consider the LP solution that sets $x_e=\frac1{2k}$ for every edge $e\in E$. It is clearly feasible for the rank
constraints (every $s-t$ path has $x$-value one). Furthermore, $x(E_i)=|E_i|/(2k)=1$ for all $i\in [k]$; i.e. the
degree constraints are also satisfied. Hence the LP relaxation is feasible.

On the other hand, consider any integral solution $I\sse E$ that has $|I\cap E_i|\le k-1$ for all $i\in [k]$. It can be
checked directly that there is an $s-t$ path using only edges $E\setminus I$. Thus any integral feasible solution $J$
must have $\max_{i\in [k]}~ |J\cap E_i|\ge k$, i.e. it violates some degree-bound by at least an additive
$k-1=\Omega(\sqrt{n})$ term.

%\vspace{-2mm}
\subsection{Algorithm for crossing lattice polyhedra satisfying monotonicity}
%\vspace{-1mm}

Given this bad integrality gap for general crossing lattice polyhedra, we are interested special cases that admit good
additive approximations. In this section we consider lattice polyhedra that satisfy the following {\em monotonicity
property}, and provide an  additive approximation.
$$(*)\quad S<T \quad\Longrightarrow\quad |\rho(S)|<|\rho(T)|,\quad \mbox{for all }S,T\in\f$$

As noted earlier, this property is satisfied by all matroids, and so our results generalize that of Kiraly et
al.~\cite{KLS08}. In the rest of this section we prove Theorem~\ref{t.BoundedLatticePolyhedra}. The algorithm is again
based on iterative relaxation. At each iteration, we maintain the following:
\begin{itemize}
%\vspace{-2mm}
\item $F\sse E$ of elements that have been chosen into the solution.
%\vspace{-2mm}
\item $E'\sse E\setminus F$ of undecided elements. \vspace{-2mm}\item $W\sse [m]$ of degree bounds.
\end{itemize}
%\vspace{-2mm}

 Initially $E'=E$, $F=\emptyset$ and $W=[m]$. In a generic iteration with $E',F,W$, we solve the following
LP relaxation on variables $\{x_e\mid e\in E'\}$, called $\lplatt(E',F,W)$:
\begin{align*}
 \min\ &c^Tx\\
  &x(\rho(S))\ge r(S)-|F\cap \rho(S)|,& \forall S\in\f \\
 &a_i-|F\cap E_i|\le x(E_i)\le b_i-|F\cap E_i|,& \forall i\in W \\
 &0\le x_e\le 1,& \forall e\in E'.
%  &x(E_i)\le b'_i &\forall i\in W\\
%  &0\le x_e\le 1 &\forall e\in E.
\end{align*}

 Consider an optimal basic feasible solution $x$ to the above LP relaxation. The algorithm does one of the
following in iteration $(E',F,W)$, until $E'=W=\emptyset$.
\begin{enumerate}
%\vspace{-2mm}
\item \label{step:lattp-0edge} If there is $e\in E'$ with $x_e=0$, then $E'\leftarrow E'\setminus \{e\}$.
%\vspace{-2mm}
\item \label{step:lattp-1edge} If there is $e\in E'$ with $x_e=1$, then $F\leftarrow F\cup \{e\}$ and
$E'\leftarrow E'\setminus \{e\}$. \vspace{-2mm} \item \label{step:lattp-deg} If there is $i\in W$ with $|E_i\cap E'|\le
2\Delta$, then $W\leftarrow W\setminus \{i\}$.
\end{enumerate}
%\vspace{-2mm}

 We note that this algorithm is a natural extension of the one for matroids~\cite{KLS08} and the one for
spanning trees~\cite{SL07}. However the correctness proof (next subsection) relies only on properties of lattice
polyhedra and the monotonicity property~$(*)$.

\subsection{Proof of Theorem~\ref{t.BoundedLatticePolyhedra}}
Assuming that one of the steps~\eqref{step:lattp-0edge}-\eqref{step:lattp-deg} applies at each iteration, it is clear
that we obtain a final solution $F^*$ that has cost at most the optimal value, satisfies the rank constraints, and
violates each degree constraint by at most an additive $2\Delta-1$. We next show that one of
~\eqref{step:lattp-0edge}-\eqref{step:lattp-deg} applies at each iteration $(E',F,W)$.
\begin{lemma} \label{lem:lat-pol-cnt} Suppose $(\f,\le)$ is a lattice satisfying the consecutive and submodular properties, and condition $(*)$, function $r$ is supermodular, and  $x$ is a
basic feasible solution to \lplatt~ with $0<x_e<1$ for all $e\in E'$. Then there exists some $i\in W$ with $|E_i\cap
E'|\le 2\Delta$.
\end{lemma}

We first establish some standard uncrossing claims (Claim~\ref{cl:latt-poly-sup} and Lemma~\ref{lem:latt-poly-unx}),
before proving this lemma. We also need some more definitions. Two elements $A,B\in\F$ are said to be {\em comparable}
if either $A\le B$ or $B\le A$; they are {\em non-comparable} otherwise. A subset $\lam\sse \f$ is called a {\em chain}
if $\lam$ contains no pair of non-comparable elements. Note that a chain in $\f$ does {\em not} necessarily correspond
to a chain in $2^E$ (with the usual subset relation) under mapping $\rho$.

Let $r'(S):=r(S)-|F\cap \rho(S)|$ for all $S\in\f$ denote the right hand side of the rank constraints in the LP solved
in a generic iteration $(E',F,W)$.
\begin{cl} \label{cl:latt-poly-sup}
$r'$ is supermodular.
\end{cl}
\begin{pf}
This follows from the consecutive and submodular properties of lattice $(\f,\le)$. Consider any $A,B\in\f$, and
\begin{eqnarray*}
|F\cap \rho_A|+|F\cap \rho_B| &=& |F\cap (\rho_A\cup \rho_B)|+|F\cap (\rho_A\cap \rho_B)|\\
&\ge &|F\cap (\rho_{A\wedge B}\cup \rho_{A\vee B})| +|F\cap (\rho_A\cap \rho_B)|\\
&\ge &|F\cap (\rho_{A\wedge B}\cup \rho_{A\vee B})| + |F\cap (\rho_{A\wedge B}\cap \rho_{A\vee B})| \\
& = & |F\cap \rho_{A\wedge B}| + |F\cap \rho_{A\vee B}|
\end{eqnarray*}
The second inequality follows from submodularity (i.e. $\rho_A\cup \rho_B\supseteq \rho_{A\wedge B}\cup \rho_{A\vee
B}$), and the third inequality uses the consecutive property $\rho_{A\wedge B}\cap \rho_{A\vee B}\sse \rho_A,~\rho_B$
(since $A\wedge B\le A,B\le A\vee B$). This combined with supermodularity of $r$ implies $r'(A)+r'(B)\le r'(A\wedge B)
+ r'(A\vee B)$ for all $A,B\in\f$.
\end{pf}

\noindent For any element $A\in\f$, let $\chi(A)\in\{0,1\}^{E'}$ be the incidence vector of $\rho(A)\sse E'$. Let
$\T:=\{A\in\f\mid x(\rho_A)=r'(A)\}$ denote the elements in $\f$ that correspond to tight rank constraints in the LP
solution $x$ of this iteration. Using the fact that $r'$ is supermodular (from above), and by standard uncrossing
arguments, we obtain the following.
\begin{lemma}\label{lem:latt-poly-unx} If $S,T\in\f$ satisfy $x(\rho_S)=r'(S)$ and $x(\rho_T)=r'(T)$, then:
$$x(\rho(S\wedge T))=r'(S\wedge T)\quad \mbox{and}\quad x(\rho(S\vee T))=r'(S\vee T)$$
Moreover, $\chi(S)+\chi(T)=\chi(S\wedge T)+\chi(S\vee T)$.
\end{lemma}
\begin{pf}
We have the following sequence of inequalities:
\begin{eqnarray*}
r'(S\wedge T) + r'(S\vee T) &\le &x(\rho_{S\wedge T})+x(\rho_{S\vee T})\\
&=& x(\rho_{S\wedge T}\cap \rho_{S\vee T}) + x(\rho_{S\wedge T}\cup \rho_{S\vee T})\\
&\le & x(\rho_{S\wedge T}\cap \rho_{S\vee T}) + x(\rho_{S}\cup \rho_{T})\\
&\le &x(\rho_{S}\cap \rho_{T}) +x(\rho_{S}\cup \rho_{T})\\
&=& x(\rho_{S}) + x(\rho_{T})\\
&=& r'(S) + r'(T)\\
&\le &r'(S\wedge T) + r'(S\vee T)
\end{eqnarray*}
The first inequality is by feasibility of $x$, the third inequality is the submodular lattice property, the fourth
inequality is by consecutive property, and the last inequality is supermodularity of $r'$. Thus we have equality
throughout, in particular $x(\rho(S\vee T))=r'(S\vee T)$ and $x(\rho(S\wedge T))=r'(S\wedge T)$. Finally since $x_e>0$
for all $e\in E'$, we also have $\chi(S)+\chi(T)=\chi(S\wedge T)+\chi(S\vee T)$.\end{pf}

Given Claim~\ref{cl:latt-poly-sup} and Lemma~\ref{lem:latt-poly-unx}, we immediately obtain the following (see
eg.~\cite{Sch03}, Chapter~60).

\begin{lemma}[\cite{Sch03}]\label{lem:latt-poly-chain}
There exists a chain $\lam\sse \T$ such that the vectors $\{\chi(A)\mid A\in\lam\}$ are linearly independent and span
$\{\chi(B)\mid B\in\T\}$.
\end{lemma}

\noindent We are now ready for the proof of Lemma~\ref{lem:lat-pol-cnt}.

\begin{pf}{\bf [Lemma~\ref{lem:lat-pol-cnt}]}
$|E'|$ is the number of non-zero variables in basic feasible $x$. Hence there exist tight linearly independent
constraints: $\lam\sse \f$ corresponding to rank-constraints and $\B\sse W$ degree-constraints, such that
$|E'|=|\lam|+|\B|$. Furthermore, by Lemma~\ref{lem:latt-poly-chain} $\lam$ is a {\em chain} in $\f$, say consisting of
the elements $S_1<S_2<\cdots <S_k$. We claim that,
\begin{equation}\label{eq:latt-cnt} |\rho(S_j)\setminus \left(\cup_{t=1}^{j-1} \rho(S_t)\right)|\ge 2,\qquad \mbox{for each }1\le j\le k\end{equation}
The above condition is clearly true for $j=1$: since $x(\rho(S_1))=r'(S_1)\ge 1$ (it is positive and integer-valued),
and $x_e<1$ for all $e\in E'$. Consider any $j\ge 2$. By the consecutive property on $S_t\le S_{j-1}< S_j$ (for any
$1\le t\le j-1$), we have $\rho(S_j)\cap \rho(S_t) \sse \rho(S_{j-1})$. So, $\rho(S_j)\setminus \left(\cup_{t=1}^{j-1}
\rho(S_t)\right)=\rho(S_j)\setminus \rho(S_{j-1})$. We now claim that $|\rho(S_j)\setminus \rho(S_{j-1})|\ge 2$, which
would prove~\eqref{eq:latt-cnt}. Since $S_{j-1}<S_j$, assumption $(*)$ implies that there is at least one element $e\in
\rho(S_j)\setminus \rho(S_{j-1})$. Moreover, if this is the only element, i.e., if $\rho(S_{j})\setminus
\rho(S_{j-1})=\{e\}$, then $\rho(S_{j-1})=\rho(S_{j})\setminus{\{e\}}$ must be true (again by property $(*)$). But this
causes a contradiction to the non-integrality of $x_e$:
$$x_e=x\left(\rho(S_{j})\right)-x\left(\rho(S_{j-1})\right)=r'\left(\rho(S_{j})\right)-r'\left(\rho(S_{j-1})\right)\in \mathbb{Z}.$$
Now, equation~\eqref{eq:latt-cnt} implies that $k=|\lam|\le \frac{|E'|}2$. Hence $|E'|\le 2|\B|$.

Suppose (for contradiction) that $|E_i\cap E'|\ge 2\Delta+1$ for all $i\in W$. Then $\sum_{i\in W} |E_i\cap E'|\ge
(2\Delta+1)\cdot |W|$. Since each element in $E'$ appears in at most $\Delta$ sets $\{E_i\}_{i\in W}$, we have
$\Delta\cdot |E'|\ge \sum_{i\in W} |E_i\cap E'|\ge (2\Delta+1)\cdot |W|$. Thus $|E'|> 2|W|\ge 2|\B|$, which contradicts
$|E'|\le 2|\B|$ from above.
\end{pf}

We are now able to prove the main result of this section:

\begin{pf}{\bf [Theorem~\ref{t.BoundedLatticePolyhedra}]}
Since the algorithm only picks $1$-elements into the solution $F$, the guarantee on cost can be easily seen. As argued
in Lemma~\ref{lem:lat-pol-cnt}, at each iteration $(E',F,W)$ one of the
Steps~\eqref{step:lattp-0edge}-\eqref{step:lattp-deg} apply. This implies that the quantity $|E'|+|W|$ decreases by 1
in each iteration; hence the algorithm terminates after at most $|E|+|I|$ iterations. To see the guarantee on degree
violation, consider any $i\in I$ and let $(E',F,W)$ denote the iteration in which it is dropped, i.e.
Step~\eqref{step:lattp-deg} applies here with $|E_i\cap E'|\le 2\Delta$ (note that there must be such an iteration,
since finally $W=\emptyset$). Since a degree bound is dropped at this iteration, we have $0<x_e<1$ for all $e\in E'$
(otherwise one of the earlier steps~\eqref{step:lattp-0edge} or~\eqref{step:lattp-1edge} applies).
\begin{enumerate}
\item {\em Lower Bound:} $a_i-|F\cap E_i|\le x(E_i\cap E')< |E'\cap E_i|\le 2\Delta$, i.e. $a_i\le |F\cap E_i| + 2\Delta-1$. The final solution contains at least all elements in $F$, so the degree lower bound on $E_i$ is violated by at
most $2\Delta-1$.
\item {\em Upper Bound:} The final solution contains at most $|F\cap E_i|+|E'\cap E_i|$ elements from $E_i$. If $E_i\cap E'=\emptyset$, the upper bound on $E_i$ is not violated. Else,  $0<x(E_i\cap E') \le b_i-|F\cap
E_i|$, i.e. $b_i\ge 1+|F\cap E_i|$, and $|F\cap E_i|+|E'\cap E_i|\le b_i+2\Delta-1$. So in either case, the final
solution violates the upper bound on $E_i$ by at most $2\Delta-1$.
\end{enumerate}
Observing that all the steps~\eqref{step:lattp-0edge}-\eqref{step:lattp-deg} preserve the feasibility of the \lplatt,
it follows that the final solution satisfies all rank constraints (since $E'=\emptyset$ finally). \end{pf}

\subsection{Algorithm for inclusion-wise ordered lattice polyhedra}
We now consider a special case of minimum crossing lattice polyhedra where the lattice $\F$ is ordered by inclusion.
I.e. the partial order in the lattice is the usual subset relation on $2^E$. This class of lattice polyhedra clearly
satisfies the monotonicity property~$(*)$, so Theorem~\ref{t.BoundedLatticePolyhedra} applies. However in this case, we
prove the following stronger guarantee for the setting with {\em only upper bounds}. This improvement comes from the
use of fractional tokens in the counting argument, as in~\cite{BKN08} (for spanning trees) and~\cite{KLS08} (for
matroids).
\begin{theorem}\label{t.InclusionwiseOrdering}
  If the underlying lattice of the minimum crossing lattice polyhedron
  problem is ordered by inclusion and only upper bounds are given, then there is an algorithm that
  computes a solution of cost at most the optimal, where all rank
  constraints are satisfied, and each degree bound is violated by at
  most an additive $\Delta-1$.
\end{theorem}

The algorithm remains the same as the one above for Theorem~\ref{t.BoundedLatticePolyhedra}. In order to prove
Theorem~\ref{t.InclusionwiseOrdering} it suffices to show the following strengthening of Lemma~\ref{lem:lat-pol-cnt}.

\begin{lemma}  Suppose $(\f,\le)$ is a lattice satisfying condition
$$S\le T \quad\Longleftrightarrow\quad \rho_S\sse \rho_T\quad \forall S,T\in \f,$$
 function $r$ is supermodular, and  $x$ is a basic feasible solution to
\lplatt~ with $0<x_e<1$ for all $e\in E'$. Then there exists some $i\in W$ with $|E_i\cap E'|\le b_i'+\Delta-1$.
\end{lemma}
\begin{pf}
The proof is very similiar to the proof of Lemma~\ref{lem:mat-intersection}. Clearly, since $\F$ is ordered by
inclusion, the consecutivity and submodularity property  are satisfied. Since $x$ is a basic feasible solution, there
exist linearly independent tight rank function- and degree bound constraints $\T$ and $\B\sse W$ such that
$$|E'|=|\T|+|\B|.$$
Using uncrossing arguments, we can assume that $(\T,\le)$ forms a chain
$$\T=\{T_1 < T_2 < \ldots < T_k\}.$$
Consider an arbitrary pair $T_i< T_{i+1}$ in $\T$, where $i\in\{1,\ldots,k-1\}$. Since $x_e>0$ for all $e\in E$ and
$\rho(T_i) \subset \rho(T_{i+1})$, it follows that $0<x(\rho(T_{i+1})\setminus{\rho(T_i)})$ and therefore, by the
integrality of $r$,
$$x(\rho(T_{i+1})\setminus{\rho(T_i)})=x(\rho(T_{i+1}))-x(\rho(T_i))=r(T_{i+1})-r(T_i)\ge 1.$$
By a similar argument, $x(\rho(T_1))\ge 1$. Thus,
$$x(E)\ge x(\rho(T_k))=\sum_{i=1}^{k-1}x(\rho(T_{i+1})\setminus{\rho(T_i)}) + x(\rho(T_1)) \ge k=|\T|$$
with equality only if $E=\rho(T_k)$. This implies that
\begin{equation}\label{eq1}
 |E'|-x(E)=|\T|+|\B|-x(E) \le |\B|.
\end{equation}
Let $E_i'=E'\cap E_i$. To prove the statement of the Lemma, it suffices to show:
$$\sum_{i\in W}(|E'_i|-b_i')=\sum_{i\in W}(|E'_i|-x(E_i))<\Delta|W|.$$
In order to prove  this, define $\Delta_e=|\{i\in W\mid e\in E_i\}|$ and consider the derivations
\begin{eqnarray*}
 \sum_{i\in W}(|E'_i|-x(E_i))&=&\sum_{i\in W}\sum_{e\in E'_i}(1-x_e)=\sum_{e\in E}\Delta_e(1-x_e)\\
 &=&\Delta\sum_{e\in E}(1-x_e)-\sum_{e\in E}(\Delta-\Delta_e)(1-x_e)\\
&\underbrace{\le}_{\mbox{eq.}(\ref{eq1})}& \Delta|\B|-\sum_{e\in E}(\Delta-\Delta_e)(1-x_e)\\
&=&\Delta |W|-\Delta |W\setminus{\B}|-\sum_{e\in E}(\Delta-\Delta_e)(1-x_e)\le \Delta|W|.
\end{eqnarray*}
Note that equality can only hold if $E=\rho(T_k)$ and $\Delta |W\setminus{\B}|+\sum_{e\in E}(\Delta-\Delta_e)(1-x_e)=
0$. The latter can only be true if $|\B|=|W|$ and $\Delta_e=\Delta$ for each $e\in E$. But this would imply that
$$\sum_{i\in \B}\chi^{E_i}=\Delta \chi^E=\Delta\chi^{T_k},$$
where $\chi^S\in \{0,1\}^{\F\times E}$ is the incidence vector of $S\in \f$ with $\chi^S_e=1$ iff $e\in \rho(S)$.
However, this  contradicts  the fact that the constraints $\T$ and $\B$ are linearly independent.
\end{pf}

}{}

\bigskip
\noindent {\bf Acknowledgement:} We thank Mohit Singh~\cite{S08} for the integrality gap for general MCST, and Chandra
Chekuri for finding an error in the arborescence result in an earlier version~\cite{BK+10} of this paper.

%%%%%%%%%%%%%%%%%%%%%%%%%%%%%%%%%%%%%%%%%%%%%%%%%%%%%%%%%%%%%%%%%%%%%%%%%%%%%%%%%%%%%%%%%%

\end{document}